\documentclass[preprint,amsmath,amssymb,superscriptaddress,showpacs]{revtex4-1}
 
\usepackage[utf8]{inputenc}
\usepackage{amsmath}
\usepackage{amssymb}
\usepackage{latexsym}
\usepackage{amsfonts}
\usepackage{epsfig}
\usepackage{psfrag}
\usepackage{graphicx}
\usepackage{hyperref}
\usepackage{color}

\definecolor{black}{rgb}{0,0,0}
\definecolor{blue}{rgb}{0,0,1}
\definecolor{green}{rgb}{0,1,0}
\definecolor{red}{rgb}{1,0,0}
\definecolor{brown}{rgb}{0.4,0.2,0}
\definecolor{darkgreen}{rgb}{0,0.7,0}

\renewcommand{\vec}[1]{\boldsymbol #1}
\newcommand{\ket}[1]{\left|#1\right>}
\newcommand{\bra}[1]{\left<#1\right|}   
\newcommand{\braket}[2]{\left<#1|#2\right>}

\newcommand{\bea}{\begin{eqnarray}}
\newcommand{\ea}{\end{eqnarray}}
\newcommand{\eea}{\end{eqnarray}}

\usepackage[percent]{overpic}
 \newlength{\imagewidth}

\begin{document}
\setlength{\imagewidth}{0.5 \linewidth}

\title{Stretching and bending dynamics in triatomic ultralong-range Rydberg molecules}
\author{Christian Fey}
\email[]{christian.fey@physnet.uni-hamburg.de }
\affiliation{Zentrum f\"ur Optische Quantentechnologien, Luruper Chaussee 149, 22761 Hamburg, Universit\"at Hamburg, Germany}
\author{Markus Kurz}
\altaffiliation[New address: ]{Institut f\"ur Physik, Albert-Einstein-Stra{\ss}e 23, 18059 Rostock, Universit\"at Rostock, Germany}
\affiliation{Zentrum f\"ur Optische Quantentechnologien, Luruper Chaussee 149, 22761 Hamburg, Universit\"at Hamburg, Germany}
\author{Peter Schmelcher}
\email[]{peter.schmelcher@physnet.uni-hamburg.de }
\affiliation{Zentrum f\"ur Optische Quantentechnologien, Luruper Chaussee 149, 22761 Hamburg, Universit\"at Hamburg, Germany}
\affiliation{The Hamburg Centre for Ultrafast Imaging, Luruper Chaussee 149, 22761 Hamburg, Universit\"at Hamburg, Germany} 

%\date{\today}

\begin{abstract}
We investigate polyatomic ultralong-range Rydberg molecules consisting of three ground state atoms bound to a Rydberg atom via $s$- and $p$-wave interactions. By employing the finite basis set representation of the unperturbed Rydberg electron Green's function we reduce the computational effort to solve the electronic problem substantially. This method is subsequently applied to determine the potential energy surfaces of triatomic systems in electronic $s$- and $p$-Rydberg states. Their molecular geometry and resulting vibrational structure are analyzed within an adiabatic approach that separates the vibrational bending and stretching dynamics. This procedure yields information on the radial and angular arrangement of the nuclei and indicates in particular that kinetic couplings between bending and stretching modes induce a linear structure in triatomic $l=0$ ultralong-range Rydberg molecules.  
\end{abstract}
%\pacs{?} 

\maketitle

%%%%%%%%%%%%%%%%%%%%%%%%%%%%%%%%%%%%%%%%%%%%%%%%%%%%%%%%%%%%%%%%%%%%%%%%%%%%%
\section{Introduction}
The attractive interaction of a highly excited electron in a Rydberg atom with a polarizable ground state atom can bind two atoms to form an ultralong-range Rydberg molecule (ULRM).
These molecules were predicted theoretically in 2000 \cite{greene_creation_2000} and were first observed in 2009 \cite{bendkowsky_observation_2009}. 
Compared to conventional molecules the most striking features of ULRM are their huge bond lengths and their oscillatory potential energy surfaces (PES) supporting a variety of equilibrium configurations.
ULRM can be divided into two different classes distinguished by the angular momentum $l$ of the unperturbed Rydberg electron: low-$l$ ULRM arising from quantum defect splitted states and high-$l$ ULRM arising from the near-degenerate manifold of hydrogenic states.
High-$l$ ULRM possess binding energies $\sim$ 1 GHz and permanent electric dipole moments in the kDebye regime while the binding energies of low-$l$ ULRM are on the order of tens of MHz and their electric dipole moments $\sim$ 1 Debye are much smaller and result from fractional high-$l$ admixtures  \cite{li_homonuclear_2011}.

Most experiments focused on the characterization of low-$l$ ULRM, see e.g. \cite{bellos_excitation_2013,anderson_photoassociation_2014,krupp_alignment_2014,sasmannshausen_experimental_2015,desalvo_ultra-long-range_2015}, but also the photoassociation of high-$l$ ``trilobite'' \cite{booth_production_2015} and recently ``butterfly'' \cite{niederprum_observation_2016} ULRM have been achieved.
Beyond the above it is very natural to consider polyatomic ULRM consisting of $N$ ground state atoms bound by the Rydberg atom. It is to be expected that polyatomic ULRM open the doorway to a plethora of novel phenomena due to the exaggerated properties of these long-ranged molecules. Indeed many of the fundamental molecular questions (molecular equilibria  and geometry, conical intersections, ultrafast decay, etc.) take now place on a completely different scale and new properties are induced according to the Rydberg character. 
Spectral signatures of trimers, tetramers and pentamers in $l=0$ Rydberg states have indeed been identified in ultra-cold dense Rb gases \cite{bendkowsky_rydberg_2010,gaj_molecular_2014}. Varying the Rydberg excitation number $n$ allowed to study the transitions from a few-body to a mean field regime where the electron interacts with up to $N\sim 10000$ ground state atoms \cite{balewski_coupling_2013}. Explaining the impact of many-body effects on the profile of the measured spectra was in the focus of subsequent theoretical works \cite{schmidt_mesoscopic_2016,schlagmuller_probing_2016}. 

Although experimentally not yet thoroughly addressed, the few existing theoretical explorations on polyatomic ULRM focused so far on high-$l$ systems. Symmetric cuts of the PES for linear ($N = 2$), triangular ($N = 3$) or quadratic ($N = 4$) configurations have been analyzed by employing symmetry adapted orbitals \cite{liu_polyatomic_2006}.
It was demonstrated that additional ground state atoms lead to a splitting of the PES which enables e.g. neon trimers to form Borromean-like states \cite{liu_ultra-long-range_2009}.
These studies have been extended recently \cite{eiles_ultracold_2016} by presenting a general formalism to determine the electronic structure of polyatomic ULRM employing hybridized diatomic orbitals and determining their PES with a focus on molecular systems with eight ground state atoms.
Triatomic high-$l$ ULRM in electric fields were very recently  investigated in \cite{fernandez_ultralong-range_2016} with a focus on the control of the electronic structure by the electric field.

Only a few of the above presented works provide a theoretical analysis of polyatomic low-$l$ ULRM \cite{schmidt_mesoscopic_2016, eiles_ultracold_2016}.
In \cite{eiles_ultracold_2016} cuts of the PES along the breathing modes for symmetric configurations of polyatomic ULRM containing 8 ground state atoms are presented and discussed. It is pointed out that the PES cuts of $l=0$ ULRM depend only weakly on the molecular geometry and that the well depth scales linearly with the number of ground state atoms whereas the PES cuts of $l=1$ and $l=2$ ULRM depend strongly on the molecular geometry. 
A discussion of the electronic and vibrational structure in the context of many-body systems is provided in \cite{schmidt_mesoscopic_2016}. However, this work does not take into account the back action on the Rydberg electron and assumes a spatially fixed Rydberg atom.         
The focus of the present article is to provide a detailed exploration of the electronic and vibrational structure of polyatomic ULRM for triatomic $l=0$ and $l=1$ systems. Hereby we aim at understanding not only the radial but also the angular configurations of the nuclei in low-$l$ states.

The paper is organized as follows. In section \ref{sec:electronic_Hamiltonian} we analyze the underlying electronic structure. In section \ref{sec:electronic_problem_approach} we set up an approach to determine the electronic structure of molecules with $N$ ground state atoms by employing the finite basis set representation of the unperturbed Rydberg electron Green's function. In section \ref{sec:PES_trimers} this approach is applied to analyze the resulting PES of low-$l$ triatomic ULRM. In section \ref{sec:vibrational_problem} we explore the vibrational dynamics of triatomic URLM. In section \ref{sec:nuc_dynamics_theoret_approach} the vibrational Hamiltonian is established and an adiabatic separation of the bending and stretching modes is accomplished.  In section \ref{sec:model} the essential properties of this Hamiltonian are analyzed for a model PES and finally, in section \ref{sec:nuclear_dynamics_ulrm} numerically obtained bending and stretching solutions for different species of triatomic ULRM are provided.

\section{Electronic structure of triatomic molecules }
\label{sec:electronic_Hamiltonian}
\subsection{Theoretical approach}
\label{sec:electronic_problem_approach}
We consider a polyatomic ultralong-range Rydberg molecule consisting of a positively charged Rydberg core, a Rydberg electron and $N$ neutral ground state atoms. The position of the electron relative to the ion core is denoted as $\vec{r}$ while the positions of the atoms relative to the ion core are denoted as $\vec{R}_j$.
In the Born-Oppenheimer approximation the electronic Hamiltonian is given by 
\begin{equation}
H^{el}_N(\vec{r})=H_0(\vec{r})+ \sum_j^N V(\vec{r},\vec{R}_j) 
\label{eqn:Hamiltonian}
\end{equation} 
where $H_0$ is the electronic Hamiltonian of the Rydberg atom and 
\begin{equation}
V(\vec{r},\vec{R}_j)=2 \pi \delta(\vec{r}-\vec{R}_j) \left( a_{s}[k(R_j)] + 3  a^{3}_{p}[k(R_j)]\overleftarrow{\nabla}_{\vec r} \cdot  \overrightarrow{\nabla}_{\vec r} \right) 
\label{eqn:pseudopotential}
\end{equation}
is a pseudopotential describing the low-energy interaction between the Rydberg electron and  the ground state atoms.
Here $a_s$ and $a_p$ denote the triplet $s$- and $p$-wave scattering lengths that depend in a semiclassical approximation on the kinetic energy of the Rydberg electron at the atomic positions $\vec{R}_j$.
For a total electronic energy $\epsilon$ this kinetic energy is determined by the wavenumber $k$ satisfying $k(R_j)=\sqrt{2\epsilon+2/R_j}$ via the relation $2 \epsilon_\text{kin}= k^2(R_j)$.

The $s$-wave pseudopotential was first developed by Fermi \cite{fermi_sopra_1934} and later extended to higher partial wave contributions \cite{omont_theory_1977}.
It has been employed successfully to describe the spectra of ULRM with and without external fields for various electronic states and different atomic species \cite{bendkowsky_observation_2009,bendkowsky_rydberg_2010,tallant_observation_2012,krupp_alignment_2014,desalvo_ultra-long-range_2015,gaj_hybridization_2015}.
Although it has been shown that a refined description of diatomic ULRM requires taking into account singlet scattering channels as well as the hyperfine structure of the ground state atoms \cite{anderson_angular-momentum_2014,anderson_photoassociation_2014, sasmannshausen_experimental_2015,bottcher_observation_2015}, we restrict our analysis to the simpler potential (\ref{eqn:pseudopotential}).
Due to the increased complexity of polyatomic systems this has been done so far in all other previous works dealing with ULRM consisting of more than one ground state atom, see e.g. \cite{bendkowsky_rydberg_2010,liu_polyatomic_2006,liu_ultra-long-range_2009,gaj_molecular_2014,schmidt_mesoscopic_2016,eiles_ultracold_2016,fernandez_ultralong-range_2016}.
We remark that the potential (\ref{eqn:pseudopotential}) yields the correct first order energy correction for spin polarized (all electronic spins are parallel) Rydberg $l=0$ states, which are in the main focus of this work.
The use of the pseudopotential within a basis set diagonalization approach is rigorously justified only when working in a limited basis set including solely states with energies sufficiently close to $\epsilon$  \cite{omont_theory_1977, fey_comparative_2015}.

We propose in this work a method equivalent to the finite basis set diagonalization of $H^\text{el}_N$ that employs Green's functions and reduces the dimensionality of the underlying eigenvalue problem.
To this aim we express the electronic wave function $\psi_\epsilon(\vec{r})$ satisfying the stationary Schr\"odinger equation $H^{el}_N \psi_\epsilon=\epsilon \psi_\epsilon$  as a solution of the Lippman-Schwinger equation
\begin{equation}
\psi_\epsilon(\vec{r})=-\int d^3r' G^0_\epsilon(\vec{r},\vec{r}') \sum_j^N V(\vec{r'},\vec{R}_j) \psi_\epsilon(\vec{r}')
\label{eqn:lipschwing}
\end{equation}
where $G_\epsilon^0(\vec{r},\vec{r}')$ is the position representation of the Green's function of $H_0$ defined as $G^0_\epsilon=(H_0-\epsilon)^{-1}$.
For a Rydberg Hamiltonian $H_0$ with eigenstates $\varphi_{nlm}$ constructed in a finite subspace $\mathcal{B}$ the
Green's function can be expressed as the sum   
\begin{equation}
G_\epsilon^0(\vec{r},\vec{r}')= \sum \limits_{\varphi_{nlm} \in \mathcal{B}} \frac{\varphi^*_{nlm}(\vec{r})\varphi_{nlm}(\vec{r}')}{\epsilon_{nl}^0-\epsilon}  \ .
\label{eqn:G_Expansion}
\end{equation}
Here $n$, $l$ and $m$ are the usual hydrogenic quantum numbers %
and $\epsilon^0_{nl}$ denotes the quantum defect energy depending on $n$ and $l$ via
$\epsilon^0_{nl}=-1/[2(n-\Delta_l)^2]$, where $\Delta_l$ is the $l$-dependent quantum defect. 
Evaluating the integral in equation (\ref{eqn:lipschwing}) with the pseudopotential (\ref{eqn:pseudopotential}) leads to
\begin{equation}
\psi_\epsilon(\vec{r})= -2 \pi \sum_{j=1}^N \left( a_{s}[k(R_j)] G_\epsilon(\vec{r},\vec{R}_j) \psi_\epsilon(\vec{R}_j)   + 3  a^{3}_{p}[k(R_j)]
\overrightarrow{\nabla}_{\vec{R}_j}G_\epsilon(\vec{r},\vec{R}_j) \cdot  \overrightarrow{\nabla}_{\vec{R}_j}\psi_\epsilon(\vec{R}_j)
  \right) \ .
  \label{eqn:lippschwing_delta}
\end{equation}
The r.h.s.\ of (\ref{eqn:lippschwing_delta}) expresses the electronic wave function as a superposition of the functions $G_\epsilon(\vec{r},\vec{R}_j)$ and $\overrightarrow{\nabla}_{\vec{R}_j} G_\epsilon(\vec{r},\vec{R}_j)$ weighted by the $4 N$ unknown coefficients $\psi_\epsilon(\vec{R}_j)$ and $\overrightarrow{\nabla}_{\vec{R}_j} \psi_\epsilon(\vec{R}_j)$.
These coefficients can be determined as solutions of a system of linear equations which is constructed by evaluating $\psi_\epsilon(\vec{r})$ in equation (\ref{eqn:lippschwing_delta}) and its gradient $\overrightarrow{\nabla} \psi_\epsilon(\vec{r})$  at the positions $\vec{R}_i$ of the $N$ ground state atoms. In a compact notation this systems reads  

\begin{equation}
\psi_\epsilon^{(\alpha)}(\vec{R}_i)=-2 \pi \sum \limits_{j=1}^N \sum \limits_{\beta=0}^3  a^{(\beta)}[k(R_j)] G_\epsilon^{(\alpha)(\beta)}(\vec{R}_i,\vec{R}_j) \psi_\epsilon^{(\beta)}(\vec{R}_j)
\label{eqn:lippschwing_algebraic}
\end{equation}
with
\begin{equation}
G_\epsilon^{(\alpha)(\beta)}(\vec{r},\vec{r}')= \sum \limits_{nlm \in \mathcal{B}} \frac{\varphi^{*(\alpha)}_{nlm}(\vec{r})\varphi^{(\beta)}_{nlm}(\vec{r}')}{\epsilon_{nl}^0-\epsilon} \ .
\label{eqn:Green_coefficients}
\end{equation}
By the Greek indices $\alpha, \beta \in \{0,1,2,3\}$ we define the four-component vectors $a^{(\alpha)}[k(R)]$, $\psi^{(\alpha)}_\epsilon(\vec{r})$ and $\varphi^{(\alpha)}_{nlm}(\vec{r})$. Their indices $\alpha \geq 1$ denote   
$a^{(\alpha \geq 1)}[k(R)]= 3 a^3_p[k(R)]$ , $\psi^{(\alpha\geq 1)}_\epsilon(\vec{r})= \nabla^{(\alpha)} \psi_\epsilon(\vec{r})$ and $\varphi^{(\alpha\geq 1)}_{nlm}(\vec{r})= \nabla^{(\alpha)} \varphi_{nlm}(\vec{r})$, where $\nabla^{(\alpha)}$ is the $\alpha$-th component of the gradient. Their indices $\alpha=0$ denote
$a^{(0)}[k(R)]= a_s[k(R)]$ , $\psi^{(0)}_\epsilon(\vec{r})= \psi_\epsilon(\vec{r})$ and $\varphi^{(0)}_{nlm}(\vec{r})= \varphi_{nlm}(\vec{r})$.
Nontrivial solutions $\psi^{(\alpha)}_\epsilon(\vec{R_i})\neq0$ of (\ref{eqn:lippschwing_algebraic}) exist only at energies $\epsilon$ where the $4N \times 4N$ matrix
\begin{equation}
M(\epsilon)_{\{\alpha,i\}\{\beta,j\}}= 2\pi a^{(\beta)}[k(R_j)] G_\epsilon^{(\alpha)(\beta)}(\vec{R}_i,\vec{R}_j) + \delta_{\alpha \beta} \delta_{i j}
\label{eqn:M_matrix}
\end{equation}
has a vanishing determinant 
\begin{equation}
\det(M(\epsilon))=0 \ .
\label{eqn:determinant}
\end{equation}
Here $\delta_{ij}$ is the Kronecker delta and the multiindices $\{\alpha,i\}$, $\{\beta,j\}$ define respectively the row and the column of the matrix $M(\epsilon)$.
Determining these energies $\epsilon$ via (\ref{eqn:determinant})  for each nuclear configuration $\vec{R}_1,\dots,\vec{R}_N$ yields the PES $\epsilon(\vec{R}_1,\dots,\vec{R}_N)$. The corresponding electronic wave function $\psi_\epsilon(\vec{r})$ can be obtained by solving the system of equations (\ref{eqn:lippschwing_algebraic}) at the energies $\epsilon(\vec{R}_1,\dots,\vec{R}_N)$ and inserting the resulting  coefficients $\psi^{(\alpha)}_\epsilon(\vec{R_i})$ into equation (\ref{eqn:lippschwing_delta}).   
The resulting PES are equivalent to the ones obtained by diagonalizing $H^{el}_N$ within the subspace $\mathcal{B}$. 
In the diatomic limit of $N=1$ and for pure $s$-wave interaction, i.e. $a^{(\alpha\geq1)}[k(R)]=0$, equation (\ref{eqn:determinant}) reduces to the well-known condition $1+2\pi a_s[k(R_1)] G_\epsilon(\vec{R}_1,\vec{R}_1)=0$, see \cite{khuskivadze_adiabatic_2002}. 

The Green's function approach can be compared to the recently proposed method employing hybridized diatomic states \cite{eiles_ultracold_2016}. At each fixed nuclear configuration both approaches reduce numerical efforts to a comparable degree by not initializing and diagonalizing the full Hamiltonian $H^{el}_N$, which is typically a dense $n_0^2\times n_0^2$ matrix, where $n_0$ denotes the quantum number of the Rydberg state of interest. In \cite{eiles_ultracold_2016} the full electronic problem is mapped to a typically $4N$ dimensional generalized eigenvalue problem by expressing the electronic wave function as a superposition of $4N$ diatomic wave functions. These diatomic wave functions need to include all eigenstates of $H^{el}_1$ within (degenerate) first order perturbation theory having eigenvalues different from the eigenvalues of the unperturbed Rydberg Hamiltonian. Consequently, an increase of the basis set size, e.g. including Rydberg states with quantum numbers $n_0+1$ and $n_0-1$, increases ultimately also the dimension of the eigenvalue problem to more than $4N$.
Contrarily, in the Green's function approach the electronic wave function (\ref{eqn:lippschwing_delta}) is, independently of the number of basis states, expressed as a superposition of $4N$ wave functions $G_\epsilon^{(0)(\beta)}(\vec{r},\vec{R}_j)$ which allows to map the electronic problem to a $4N$ dimensional linear algebraic system. However, this linear system is energy dependent and needs to be solved typically by application of root finding algorithms, cf. equation (\ref{eqn:determinant}).
One advantage of the Green's function approach for future applications is that additional interactions, like external fields, can be absorbed into the Green's function and do not increase the dimension of the system of linear equations (\ref{eqn:lippschwing_algebraic}). E.g. for electric fields one would replace the states $\varphi_{nlm}$ in (\ref{eqn:G_Expansion}) by Rydberg-Stark states. In contrast, it is not obvious how to deal with these additional interactions in \cite{eiles_ultracold_2016} without increasing the size of the required diatomic wave functions.

\subsection{Potential energy surfaces of trimers}
\label{sec:PES_trimers}
%%%%%%%%%%%%%%%%%%%%%%%%%%%%%%%%%%%%%%%%%%%%%%%%%%%%%%%%%%%%%%%%%%%%%%%%
In the following we investigate the PES of triatomic ULRM (N=2). Their PES depend in general on three internal coordinates: the internuclear separations $R_1$ and $R_2$ as well as the angle $\theta$ enclosed by $\vec{R}_1$ and $\vec{R}_2$.
To begin with, we determine the PES in first order perturbation theory by restricting the subspace $\mathcal{B}$ to a manifold of energetically degenerate eigenstates having a certain energy $\epsilon^0$ and approximating $k(R_j)\approx \sqrt{2\epsilon^0+2/R_j}$.    
For $N=2$ and pure $s$-wave interaction equation (\ref{eqn:determinant}) can be solved analytically and yields the two PES
\begin{equation}
\epsilon_\pm (\vec{R}_1,\vec{R}_2)= \epsilon^0+\frac{2 \pi a_1 g_{11}+ 2\pi a_2 g_{22} }{2} \pm \frac{1}{2} \sqrt{\left(2 \pi a_1 g_{11}- 2\pi a_2 g_{22} \right)^2 +16 \pi^2 a_1 a_2 g_{12}g_{21}}
\label{eqn:energy_first_order}
\end{equation}
with
$a_i=a_{s}[k(R_i)]$ and
\begin{equation}
g_{ij}= \sum \limits_{\overset{nlm}{\epsilon^0_{nl}=\epsilon^0} } \varphi^{*}_{nlm}(\vec{R}_i)\varphi_{nlm}(\vec{R}_j) \ .
\end{equation}
This expression was also obtained in \cite{liu_ultra-long-range_2009} for high-$l$ states by using hybridized diatomic orbitals. The sum of the two solutions yields, except for the constant offset $2\epsilon^0$,  exactly the sum of the diatomic PES which means illustratively that at each point $(\vec{R_1},\vec{R}_2)$ a total energy $2 \pi a_1 g_{11}+ 2\pi a_2 g_{22}$ is distributed to an upper and a lower PES $\epsilon_+$ and $\epsilon_-$ accordingly to the expression in (\ref{eqn:energy_first_order}).
\subsubsection{Electronic $s$-Rydberg states}
The main focus of this work lies on $l=0$ ULRM. To obtain the first order energy for a state with quantum number $n$ from equation (\ref{eqn:energy_first_order}) we employ
$\epsilon^0=\epsilon^0_{n,l=0}$ as well as $g_{ij} =\varphi_{n,0,0}(R_i) \varphi_{n,0,0}(R_j) $ and the only nonzero solution of equation (\ref{eqn:energy_first_order}) reduces to 
\begin{equation}
\epsilon= \epsilon^0_{n,0} +2 \pi a_s[k(R_1)] \varphi^2_{n,0,0}(R_1)
+2 \pi a_s[k(R_2)] \varphi^2_{n,0,0}(R_2) 
\label{eqn:energy_first_order_s}
\end{equation}
which corresponds consequently to the sum of the diatomic PES.
This result can be derived equivalently by evaluating the pseudopotential (\ref{eqn:pseudopotential}) for the isotropic wave function $\varphi_{n,0,0}(\vec{r})$.
Similarly, the PES for a $l=0$ triatomic ULRM with $s$- and additional $p$-wave interaction can be derived to
\begin{align}
\epsilon=& \epsilon^0_{n,l=0} +2 \pi a_s[k(R_1)] \varphi^2_{n,0,0}(R_1)
+2 \pi a_s[k(R_2)] \varphi^2_{n,0,0}(R_2) \nonumber \\
&+ 6 \pi a^{3}_{p}[k(R_1)]|\overrightarrow{\nabla}\varphi_{n,0,0}(R_1)|^2
+ 6 \pi a^{3}_{p}[k(R_2)]|\overrightarrow{\nabla}\varphi_{n,0,0}(R_2)|^2 \ .
\label{eqn:energy_first_order_sp}
\end{align}

Beyond first order perturbation theory the pseudopotential couples the $l=0$ states to the energetically adjacent hydrogenic high-$l$ states. E.g. this coupling causes the small permanent dipole moment in diatomic $l=0$ Rb and Cs ULRM \cite{li_homonuclear_2011}.

In the following we study these high-$l$ admixtures for a $^{87}\text{Rb}$ ULRM in a 43$s$ Rydberg state. Due to its quantum defect this state lies approximately 13 GHz below the manifold of the $n=40$, $l>2$ states and approximately 93 GHz above the manifold of the $n=39,l>2$ states. According to the denominator in the Green's function expansion (\ref{eqn:G_Expansion}) the most significant admixtures will stem from the $n=40$ manifold while the impact of other manifolds will be energetically suppressed. However, there is no convergence of the PES when increasing the number of basis states \cite{fey_comparative_2015} and the pseudopotential (\ref{eqn:pseudopotential}) was only derived for a finite basis set of energetically degenerate or quasi-degenerate states \cite{omont_theory_1977}. Therefore we include just the $n=40$ manifold and model the system by the Green's function
 
\begin{equation}
G_\epsilon^{(\alpha)(\beta)}(\vec{r},\vec{r}')= \frac{\varphi^{*(\alpha)}_{43,0,0}(\vec{r})\varphi^{(\beta)}_{43,0,0}(\vec{r}')}{\epsilon_{43,0}^0-\epsilon} + \sum \limits_{l\geq 3,m} \frac{\varphi^{*(\alpha)}_{40,lm}(\vec{r})\varphi^{(\beta)}_{40,lm}(\vec{r}')}{\epsilon_{40,l}^0-\epsilon} \ ,
\label{eqn:Green_33}
\end{equation}
where $\varphi_{n,l,m}$ are phase shifted Coulomb wave functions taking into account quantum defects $\Delta_0=3.13$ and $\Delta_{l>2}=0$. 
The energy dependence of the triplet scattering lengths is included by approximating  $k(R_j)\approx \sqrt{2\epsilon^0_{43,0}+2/R_j}$ and employing the phase shift data as previously used and presented in \cite{kurz_electrically_2013}.
The PES are then obtained via equation (\ref{eqn:determinant}).

%%%%%%%%%%%%%%%%%%%%%%%%%%%%%%%%%%%%%%%%%%%%%%%%%%%%%%%%%%%%%%%%%%%%%%
\begin{figure*}[h]
\includegraphics[width=\linewidth]{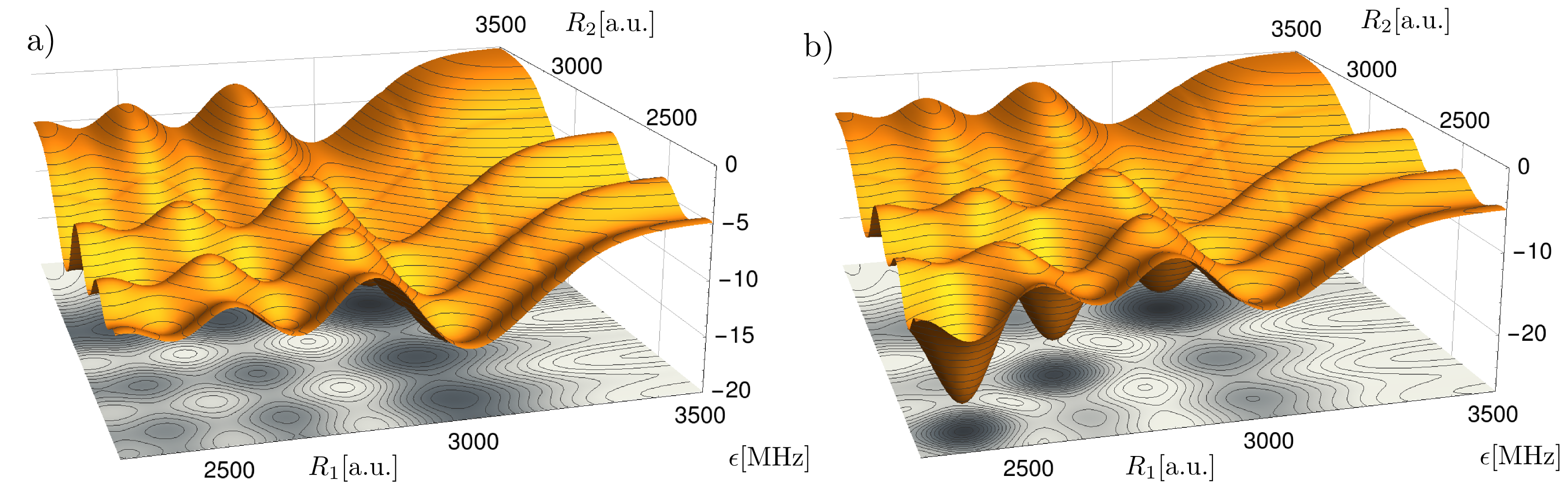}
\caption{Potential energy surface $\epsilon(R_1,R_2,\theta)$ for the Rb $43s$ trimer as a function of the internuclear separations $R_1$ and $R_2$ at fixed angles a) $\theta=\pi$ and b) $\theta=0$. The zero energy has been set to the energy of the $43 s$ Rydberg state.}
\label{fig:PES_rb40s}
\end{figure*}
%%%%%%%%%%%%%%%%%%%%%%%%%%%%%%%%%%%%%%%%%%%%%%%%%%%%%%%%%%%%%%%%%%
\begin{figure}[h]
\includegraphics[width=1.0 \imagewidth]{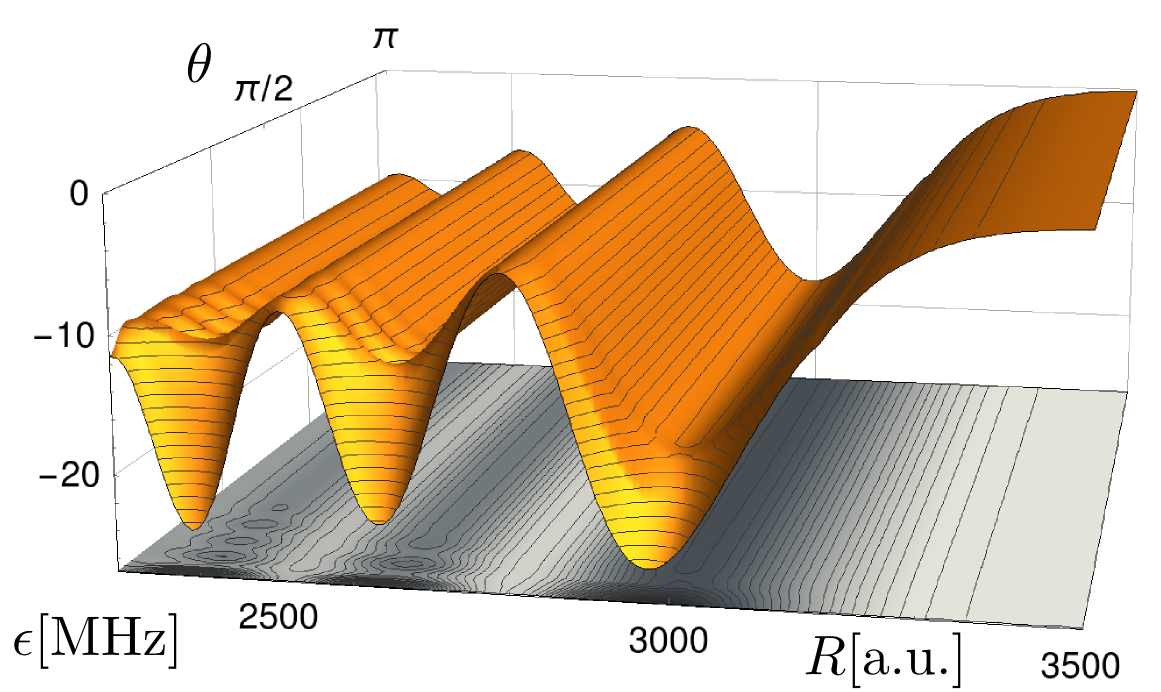}
\caption{Potential energy surface $\epsilon(R_1,R_2,\theta)$ for the 43s trimer as a function of the internuclear separations $R_1 =R_2=R$ and the angle $\theta$. The zero energy has been set to the energy of the $43 s$ Rydberg state.}
\label{fig:PES_rb40_rt}
\end{figure}
%%%%%%%%%%%%%%%%%%%%%%%%%%%%%%%%%%%%%%%%%%%%%%%%%%%%%%%%%%%%%%
 We present cuts of the PES for fixed angles $\theta=\pi$ and $\theta=0$ in Fig.\ \ref{fig:PES_rb40s} and for variable $\theta$ but constrained separations $R_1=R_2$ in Fig.\ \ref{fig:PES_rb40_rt}.
Conveniently, the energy is expressed as an equivalent frequency detuning which is simply the energy divided by the Planck constant $h$.
While the $s$-wave interaction dominates at large bond lengths (low kinetic energy of the electron close to the classical turning point), the impact of the $p$-wave interaction grows with decreasing internuclear separations which leads finally to an avoided crossing between the $l=0$ and the above lying high-$l$ states at $R\approx 1400$ a.u.\ (not visible in Fig.\ \ref{fig:PES_rb40s}).
To analyze the topology and angular dependence of the PES in regions of large bond length we focus therefore on effects of the $s$-wave interaction. The PES can be compared to the first order approximation for pure $s$-wave interaction in equation (\ref{eqn:energy_first_order_s}).
In this limit the adiabatic energy surface of the $l=0$ trimer is separable and isotropic, i.e. independent of the angle $\theta$. This approximation describes the shape of the potential in Fig.\ \ref{fig:PES_rb40s} a) qualitatively well.
The surface possesses a pronounced 18 MHz deep minimum at $R_1 = R_2 \approx 3000$ a.u.\ corresponding to the position of the outer maximum of the radial electronic wave function as well as several 10-15 MHz deep minima corresponding to combinations of other maxima in the electronic wave function. 

Corrections to the first order approximation become important when contributions of high angular momentum wave functions present in (\ref{eqn:Green_33}) are non-negligible.
The character of anisotropy induced by the $s$-wave interaction can be understood by analyzing the off-diagonal elements of $M(\epsilon)_{\{0,i\}\{0,j\}}$ with $i\neq j$. They contain the coefficients $G_\epsilon^{(0)(0)}(\vec{R}_1,\vec{R}_2)$, which, for the particular Green's function (\ref{eqn:Green_33}), are the only terms in $M(\epsilon)_{\{0,i\}\{0,j\}}$ that depend explicitly on the angle $\theta$. The sum including the high angular momentum wave functions in $G_\epsilon^{(0)(0)}(\vec{R}_1,\vec{R}_2)$ is proportional to the electronic trilobite wave function \cite{greene_creation_2000} with one ground state atom at position $\vec{R}_1$ evaluated at position $\vec{R}_2$.
The magnitude of this term is drastically increased in regions where $\vec{R}_1 \approx \vec{R}_2$ which consequently affects the shape of the energy surface in this part of configuration space.
This correlation is clearly visible 
in Fig.\ \ref{fig:PES_rb40s} b) by the pronounced peaks on the $R_1=R_2$ diagonal and in Fig.\ \ref{fig:PES_rb40_rt} where the PES depends only very weakly on $\theta$ except for regions where $\theta \approx 0$.
These results suggests that the observed triatomic ULRM \cite{bendkowsky_rydberg_2010,gaj_molecular_2014} having approximately twice the binding energy of the diatomic states correspond to nuclear configurations with $\theta > 0$ while there might also exist triatomic states with $\theta\approx0$ having deeper binding energies. However, 
the analysis of these configurations is not part of this work as an accurate description would probably require to take into account additional interactions between the ground state atoms.

\subsubsection{Electronic $p$-Rydberg states}
\begin{figure*}[h]
\includegraphics[width=0.99\linewidth]{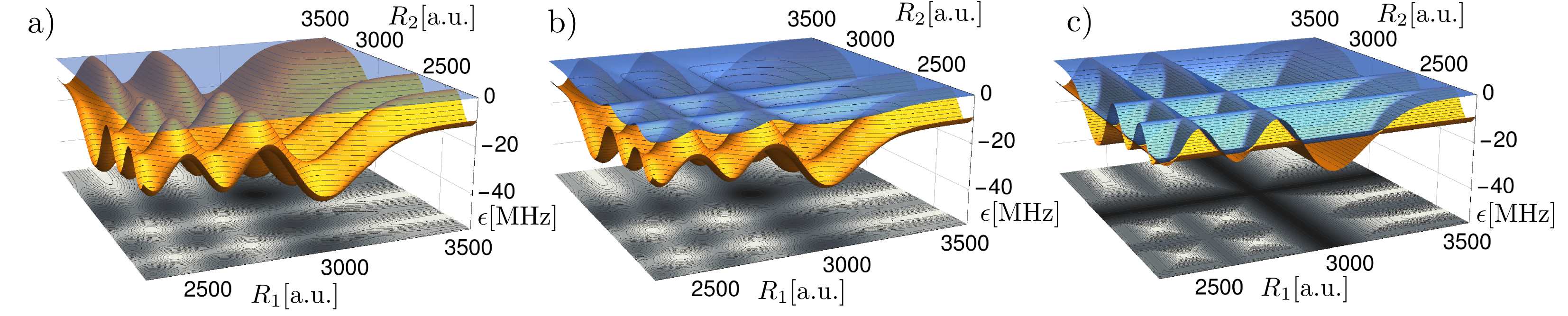}
\caption{Potential energy surfaces $\epsilon_+$ (transparent blue) and $\epsilon_-$ (yellow) in first order perturbation theory  for the triatomic $42p$ state with pure $s$-wave interaction as a function of the internuclear separations $R_1$ and $R_2$ at fixed angles a) $\theta=\pi$, b) $\theta=3\pi/4$ and c) $\theta=\pi/2$. The zero energy has been set to the energy of the $42 p$ Rydberg state.}
\label{fig:PES_rb40p}
\end{figure*}

\begin{figure}[h]
\includegraphics[width=0.99 \imagewidth]{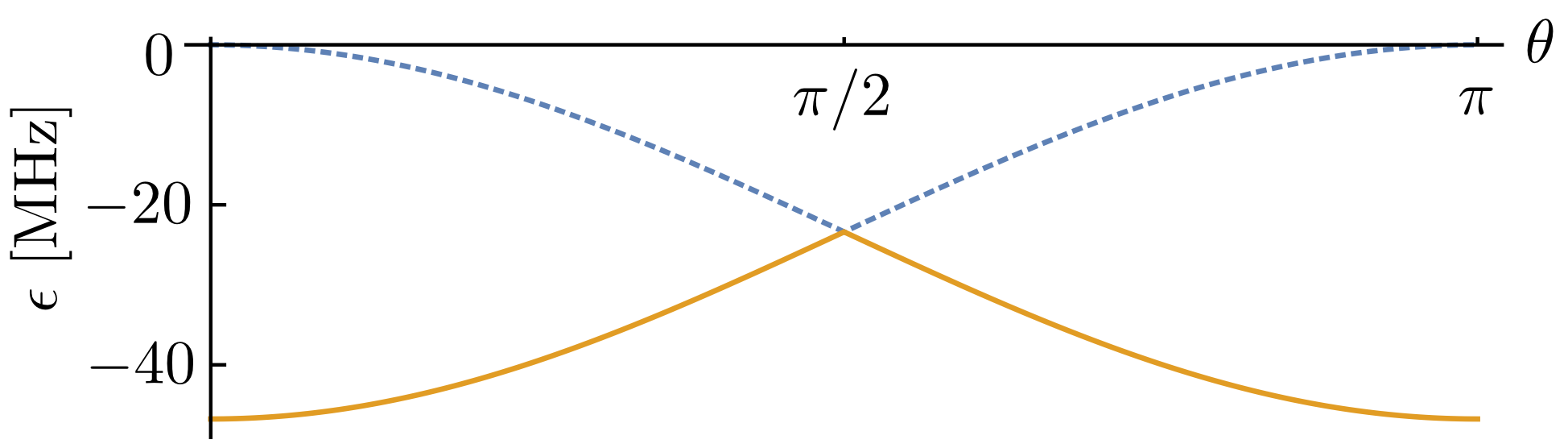}
\caption{Potential energy surfaces $\epsilon_+$ (blue) and $\epsilon_-$ (yellow) for the triatomic $42p$ state with pure $s$-wave interaction as a function of the angle $\theta$ at fixed internuclear distances $R_1 =R_2= 2930 \text{ a.u.}$. The zero energy has been set to the energy of the $42p$ Rydberg state.}
\label{fig:PES_rb40p_theta}
\end{figure}

To illustrate qualitative changes in the electronic structure when going to higher angular momentum numbers $l$ we present the PES of a triatomic $^{87}\text{Rb}$ system in a $42p$ state obtained with the potential (\ref{eqn:pseudopotential}) in the limit of pure $s$-wave interaction and in first order perturbation theory.
As electronic wave function $\varphi_{42,1,m}$ we use a phase shifted Coulomb wave function taking into account a $\Delta_{1}=2.65$ quantum defect.
The two resulting PES can directly be calculated from equation (\ref{eqn:energy_first_order}).
In Fig.\ \ref{fig:PES_rb40p} we depict the radial dependence of the PES for fixed angles between $\pi$ and $\pi/2$. 
In contrast to the triatomic $l=0$ states, the first order PES possesses already an angular dependence.
For $\theta=\pi$ the upper PES $\epsilon_+$ is zero while the lower PES $\epsilon_-$ equals the sum of the diatomic PES.
At this angle it resembles therefore the previously discussed PES of triatomic $l=0$ state and possesses a pronounced 47 MHZ deep miminum at $R_1 =R_2\approx 2930$ a.u.. 
For smaller angles $\epsilon_-$ flattens while $\epsilon_+$ deepens until, at $\theta=\pi/2$, the two PES touch along several curves, e.g. along the diagonal $R_1=R_2$.
At this particular angle the resulting topology of the two PES in Fig.\ \ref{fig:PES_rb40p} c) can be interpreted geometrically as two intersecting purely diatomic PES $\epsilon_1(R_1)= \epsilon^0_{42,1} +2 \pi a_1 g_{11}$ and $\epsilon_2(R_2)= \epsilon^0_{42,1} + 2\pi a_2 g_{22}$.
The structure of the PES for angles $\theta<\pi/2$ can be deduced from Fig.\ \ref{fig:PES_rb40p} since the PES are symmetric with respect to reflections of $\theta$ around $\theta=\pi/2$ which is a consequence of the symmetry of the involved $l=1$ electronic wave functions.
This symmetry is also visible in Fig.\ \ref{fig:PES_rb40p_theta} showing the angular dependence of the PES for a situation where the two ground state atoms are fixed at the outer well $R_1=R_2=2930$ a.u..
The lower curve possesses two equilibrium positions at $\theta=0$ and $\theta=\pi$ and intersects with the upper curve at $\theta=\pi/2$.
%%%%%%%%%%%%%%%%%%%%%%%%%%%%%%%%%%%%%%%%%%%%%%

In the following analysis of the vibrational structure we will exclusively focus on states bound in the $\theta=\pi$ well where the dominant angular dependence of the PES is already given by the first order result and does not change qualitatively when including higher order corrections. Therefore we will not discuss the effects of high-$l$ admixtures for this system.

%%%%%%%%%%%%%%%%%%%%%%%%%%%%%%%%%%%%%%%%%%%%%%%%%%%%%%%%%%%%%%%%%%%%
\section{Vibrational Dynamics of Trimers}
\label{sec:vibrational_problem}
\subsection{Theoretical approach}
\label{sec:nuc_dynamics_theoret_approach}
After having analyzed the electronic structure of triatomic ULRM in $s$- and $p$-states we now focus on the vibrational structure.
The degrees of freedom of the 9 dimensional full nuclear Hamiltonian can be reduced by separating the center of mass motion such that the remaining Hamiltonian depends only on coordinates relative to the ion core and reads

\begin{equation}
H_N^{rel}= \frac{1}{m}\left(\vec{P}_{1}^2+\vec{P}_{2}^2 +\vec{P}_{1} \cdot \vec{P}_{2} \right) + \epsilon(\vec{R}_{1},\vec{R}_{2}) \ .
\label{eqn:Hamiltonian_nuc_full}
\end{equation}
This Hamiltonian is equivalent to the one given in equation (16) of \cite{liu_ultra-long-range_2009} expressed in Jacobi coordinates, which was employed to characterize the nuclear wave function of triatomic ultralong-range Rydberg molecules in high-$l$ states via a consequently performed normal mode analysis. $H^\text{rel}_N$ can rigorously be separated into a purely vibrational part, depending solely on the internal coordinates $R_1$, $R_2$, $\theta$ and a rotational-vibrational part, i.e. $H_N^{rel}=H^\text{vib}+H^\text{rovib}$. The vibrational part reads \cite{handy_variational_1982,handy_derivation_1987} 
\begin{align}
H^\text{vib}&= \frac{1}{m}\left[-\frac{\partial^2}{\partial R_1^2}-\frac{\partial^2}{\partial R_2^2}- \cos \theta \frac{\partial}{\partial R_1}\frac{\partial}{\partial R_2}  \right] \nonumber \\
&- \frac{1}{m} \left(\frac{1}{R_1^2}+ \frac{1}{R_2^2}- \frac{\cos\theta}{R_1 R_2} \right)\left(\frac{\partial^2}{\partial \theta^2} + \cot \theta \frac{\partial}{\partial \theta}  \right) \nonumber \\
&-\frac{1}{m}\left(\frac{1}{R_1 R_2}-\frac{1}{R_2}\frac{\partial}{\partial R_1}
-\frac{1}{R_1}\frac{\partial}{ \partial R_2}
\right)\left(\cos \theta + \sin \theta \frac{\partial}{\partial \theta} \right) \nonumber\\
&+ \epsilon(R_1,R_2,\theta) \ ,
\label{eqn:Hamiltonian_nuclear}
\end{align}
where the volume element to evaluate matrix elements of this Hamiltonian is given by $dR_1 dR_2 d\theta \sin(\theta)$. 
The total angular momentum $\vec{L}$ of the system is conserved and here we focus on the case $\vec{L}=0$ for which $H^\text{rovib}=0$.

\subsubsection{Separation of stretching and bending dynamics }
To find eigenfunctions $\chi_\nu(R_1,R_2,\theta)$ of the vibrational Hamiltonian one can in a first step diagonalize the Hamiltonian
\begin{equation}
H^\text{str}(R_1,R_2;\theta)=  \frac{1}{m}\left[-\frac{\partial^2}{\partial R_1^2}-\frac{\partial^2}{\partial R_1^2}- \cos \theta \frac{\partial}{\partial R_1}\frac{\partial}{\partial R_2}  \right] + \epsilon(R_1,R_2,\theta) 
\label{eqn:Hamiltonian_strechting}
\end{equation}
which we denote as the stretching Hamiltonian as it depends only parametrically on $\theta$. Its eigenfunctions  $\chi_j^\text{str} (R_1,R_2;\theta)$ with energy curves $E^\text{str}_j(\theta)$ describe the stretching dynamics in the coordinates $R_1$ and $R_2$.
They permit to expand the eigenfunctions $\chi_\nu(R_1,R_2,\theta)$ of the full vibrational Hamiltonian $H^\text{vib}$ with energy $E_\nu$ as 
\begin{equation}
\chi_\nu(R_1,R_2,\theta)=\sum_j \chi_j^\text{str} (R_1,R_2;\theta) \chi_{j \nu} ^\text{ben}(\theta)  \ .
\label{eqn:adiabatic_expansion}
\end{equation}
where $\chi_{j \nu} ^\text{ben}(\theta)$ describes the bending dynamics. 
Projecting the full Schr\"odinger equation $H^\text{vib} \chi_\nu =E_\nu \chi_\nu$ onto the stretching solutions $\chi^\text{str}_j$ leads to coupled channel equations. Their diagonalization is numerically involved and leads ultimately to the exact solutions $\chi_\nu(R_1,R_2,\theta)$.   

Approximate solutions can be found by following an adiabatic approach similar to the rigid rotor approximation \cite{gonzalez-ferez_rovibrational_2005}. Conditions under which this approximation (\ref{eqn:bending_mode}) becomes valid are provided and discussed in detail in the appendix \ref{sec:adiabatic}. This approach is based on the assumption that the typical internuclear separations $\left<R_1\right>$ and $\left<R_2\right>$ are large for an ULRM and imply therefore large moments of inertia for the bending motion such that the dynamics in the bending mode is slow compared to the stretching dynamics. In this case the coupled channel equations decouple adiabatically and the bending modes of the  $j$-th stretching mode can be described by an effective Schr\"odinger equation
with the bending Hamiltonian
\begin{align}
H_j^\text{ben}=&- I_j \left(\frac{\partial^2}{\partial \theta^2} + \cot \theta \frac{\partial}{\partial \theta}  \right) \nonumber \\
&-J_j \left(\cos \theta + \sin \theta \frac{\partial}{\partial \theta} \right)
+ E^\text{str}_j(\theta) \ ,
\label{eqn:bending_mode}
\end{align}
where
\begin{equation}
I_j= \frac{1}{m} \left( \left<\frac{1}{R_1^2}\right>_j+ \left<\frac{1}{R_2^2} \right>_j- \left<\frac{1}{R_1 R_2}\right>_j \cos \theta \right)
\label{eqn:moment_of_inertia}
\end{equation}
and
\begin{equation}
J_j= \frac{1}{m} \left<\frac{1}{R_1 R_2}\right>_j
.
\end{equation}
By $\left<\cdot\right>_j$ we denote the $\theta$-dependent expectation values with respect to $\chi_j^\text{str}$, e.g. $\left<1/R_1R_2\right>_j=\int d R_1 d R_2 |\chi_j^\text{str}(R_1,R_2;\theta)|^2  1/R_1R_2$. Semiclassically $I_j$ can be interpreted as the inverse of the moment of inertia for the bending motion. In the next sections this approach will be employed to investigate the nuclear dynamics of triatomic ultralong-range Rydberg molecules in electronic $s$- and $p$-states.
\subsubsection{Multiconfiguration Time-Dependent Hartree method}

A powerful method to obtain numerically exact eigenstates of the vibrational Hamiltonian $H^\text{vib}$ given in (\ref{eqn:Hamiltonian_nuclear}) is to employ the improved relaxation scheme of the Multiconfiguration Time-Dependent Hartree (MCTDH) package
\cite{MEYER1990,Beck20001,meyer2003,meyer_calculation_2006,meyer_2008,
meyer_2012,mctdh:package}. In the following we briefly describe this method while more comprehensive introductions are given in \cite{meyer_calculation_2006,meyer_2012}.

Originally MCTDH was developed as a tool for propagating wave packets in high dimensional spaces. Within this approach any time-dependent nuclear wave function $\chi(R_1,R_2,\theta,t)$ of the triatomic ULRM is expressed as
\begin{equation}
\chi(R_1,R_2,\theta,t)=\sum_{i_1=1}^{n_1} \sum_{i_2=1}^{n_2}\sum_{i_3=1}^{n_3} A_{i_1,i_2,i_3}(t) \phi^{(1)}_{i_1}(R_1,t) \phi^{(2)}_{i_2}(R_2,t)  \phi^{(3)}_{i_3}(\theta,t) \ ,
\label{eqn:mctdh_expansion}
\end{equation}   
with $A_{i_1,i_2,i_3}(t)$ being a time-dependent coefficient, $\phi_{i_d}^{(d)}$ the so-called $i_d$-th single particle function of the $d$-th degree of freedom and
 $n_d$ the number of single particle functions employed for the $d$-th degree of freedom. By introducing a multiindex $I$ equation (\ref{eqn:mctdh_expansion}) can be written compactly as $\chi(R_1,R_2,\theta,t)=\sum_I A_I \phi_I $.
The key idea of the MCTDH propagation algorithm consist of keeping the number of necessary single particle functions small by employing variationally optimized $A_I$ and $\phi_I$ which are generated from an initial state by the MCTDH equations of motion.
 
By performing imaginary time propagation the MCTDH equations of motion allow also to determine the ground state of $H^\text{vib}$. For excited eigenstates an algorithm called improved relaxation can be derived by varying the energy functional $\bra{\chi}H^\text{vib} \ket{\chi}$ with respect to $A_I$ and $\phi_I$ under the additional constraints $\sum_IA^*_I A_I=1$ and $\left<\phi^{(d)}_i|\phi^{(d)}_j\right>=\delta_{ij}$, which ensure the normalization of $\chi$ and the orthonormality of the single particle functions.
In the sequence of this algorithm an initial state is propagated step-wise by determining an eigenvector $A_I$ of the Hamiltonian matrix $\bra{\phi_I}H^\text{vib}\ket{\phi_{J}}$ given in the instantaneous basis $\phi_I$ and relaxing subsequently the single particle orbitals by imaginary time propagation of the MCTDH equations of motion for the $\phi^{(d)}_{i_d}$ while keeping the coefficients $A_I$ entering these equations fixed. This procedure is repeated till $\chi$ converges to a stationary solution of $H^\text{vib}$. By following different eigenvectors $A_I$ one is able to determine not only the ground state but also vibrationally excited states. 

In order to compare our results obtained by the adiabatic separation of stretching and bending motion to the MCTDH results we employ in a first step the improved relaxation in block form, see \cite{meyer_2012}, yielding approximate results for typically 40 of the energetically lowest eigenstates. In a second step we select out of this spectrum the relevant states corresponding to our adiabatic solutions and relax them individually by the improved relaxation scheme. Typically $n_1=n_2=n_3=10$ single particle functions for each degree of freedom are sufficient to ensure convergence on the relevant energy scales.

\subsection{Model potential energy surface}
\label{sec:model}
In order to develop a basic understanding of the properties of the underlying vibrational Hamiltonian, to quantify the influence of different characteristics of the PES on the vibrational structure and to explicate how these features are captured by the adiabatic approach, we will first of all analyze the nuclear motion for a simple model PES 
\begin{equation}
\epsilon(R_1,R_2)= \frac{1}{2}m \omega^2 \left( \left(R_1-l_0\right)^2+ \left(R_2-l_0\right)^2 \right) .
\label{eqn:model_pot}
\end{equation}
Here $\omega$ describes the strength of the potential along the radial directions whereas the parameter $l_0$ describes the bond length of the molecule.
Equation (\ref{eqn:model_pot}) can be interpreted as a Taylor expansion of the potential energy surfaces discussed in section \ref{sec:PES_trimers} around the outer potential well in the approximately isotropic regions.

For the model potential the stretching Hamiltonian $H^\text{str}$ in equation (\ref{eqn:Hamiltonian_strechting}) can be diagonalized analytically. The transformations
$R_+=(R_1+R_2)/2+l_0$ and $R_-=(R_1-R_2)/2$ lead to the harmonic oscillator Hamiltonian
\begin{equation}
H^\text{str}= -\frac{1}{2 m_+} \frac{\partial^2}{\partial R^2_+} - \frac{1}{2 m_-} \frac{\partial^2}{\partial R^2_-}  +\omega^2 \left( R_+^2 + R_-^2 \right)
\label{eqn:Hamiltonian_toy}
\end{equation}
with the $\theta$-dependent effective masses $m_+=1/(1+ \frac{\cos\theta}{2})$ and $m_-=1/(1- \frac{\cos\theta}{2})$. Here the coordinates $R_+$ and $R_-$ describe oscillations in the symmetric and antisymmetric stretching modes. The spectrum of $H^\text{str}$ reads  
\begin{equation}
E_{n_+ n_-}^\text{str}(\theta)=\omega\left(\sqrt{2+\cos \theta} \left(n_+ +\frac{1}{2} \right) +\sqrt{2-\cos \theta} \left(n_- +\frac{1}{2} \right) \right) \ ,
\label{eqn:energy_toy}
\end{equation}
where $n_+$ and $n_-$ are quantum numbers counting the nodes of the stretching functions $\chi^\text{str}_{n_+ n_-}(R_1,R_2;\theta)$ along the $R_+$ and $R_-$ direction. The parity of $n_-$ determines the exchange symmetry with respect to permutations of the ground state atom positions $R_1$ and $R_2$. Therefore even $n_-$ imply bosonic, while odd $n_-$ imply fermionic states. 

\begin{figure}[h]
\includegraphics[width=\imagewidth]{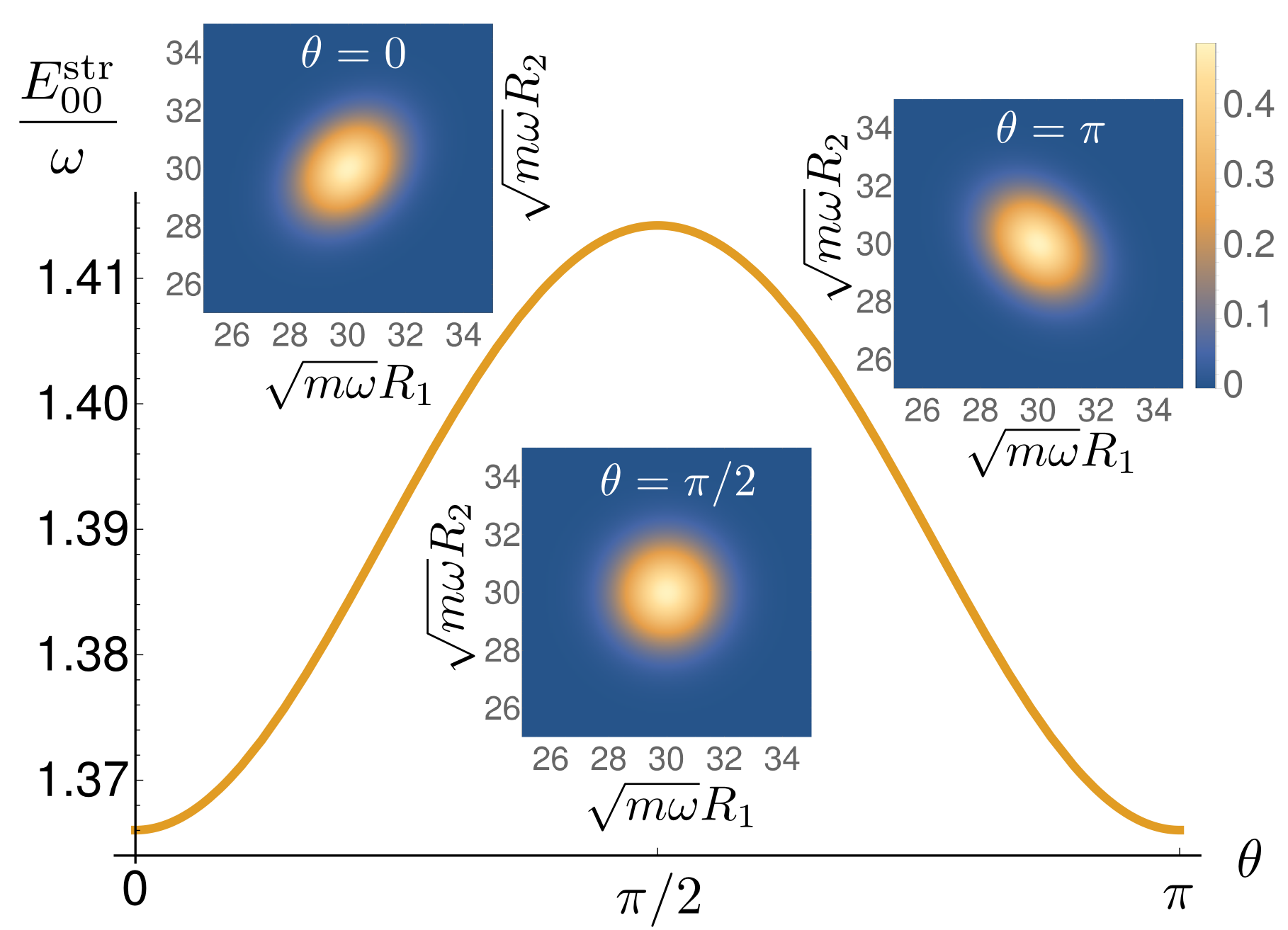}
\caption{Adiabatic energy curve $E^\text{str}_{00}(\theta)$ of the lowest stretching mode as a function of the angle $\theta$. The insets depict the stretching function $\chi^\text{str}_{00}(R_1,R_2;\theta)$ at particular angles $\theta$ for the parameter $\sqrt{m \omega}l_0=30$.}
\label{fig:00}
\end{figure}

\begin{figure}[h]
\includegraphics[width=\imagewidth]{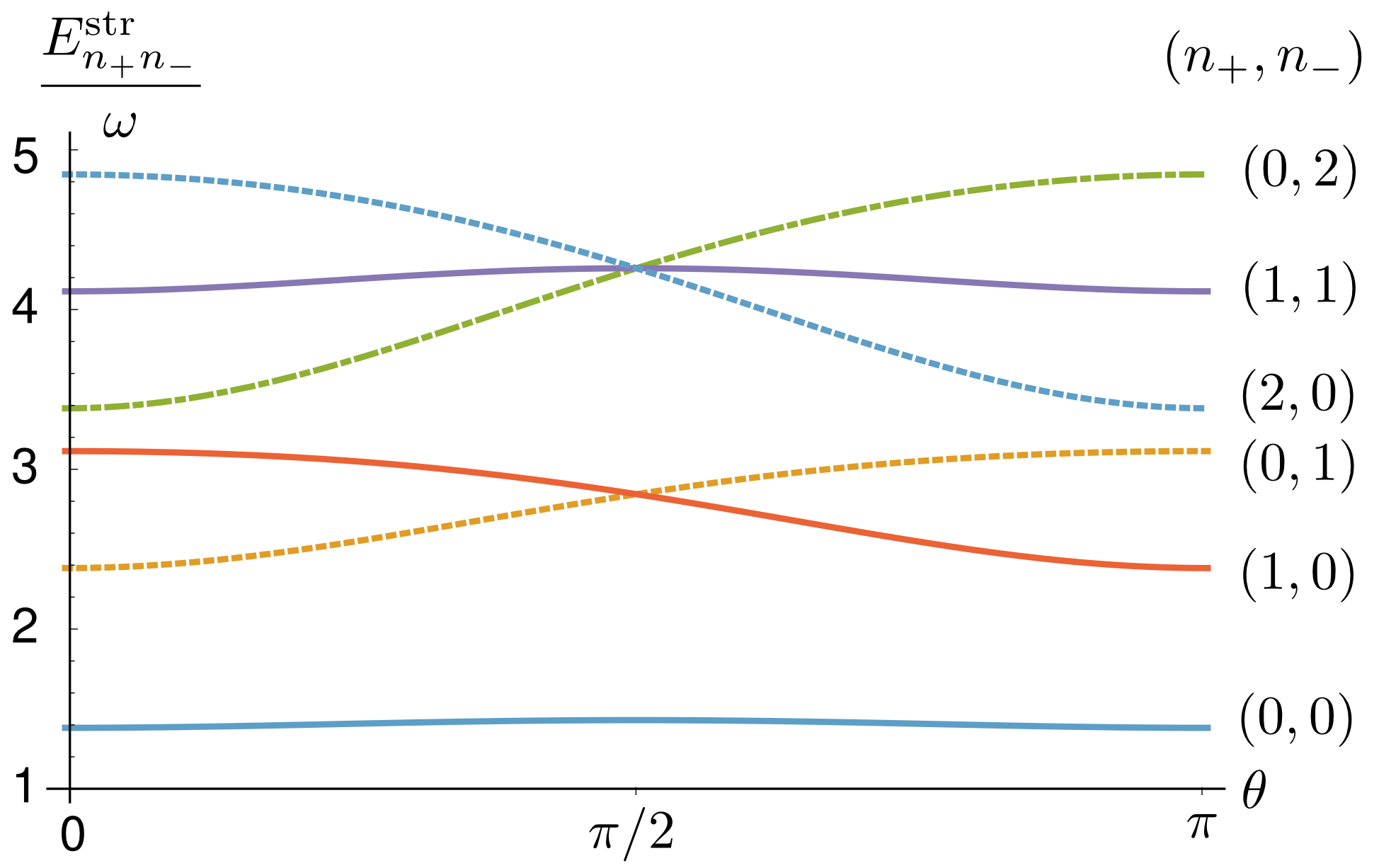}
\caption{Adiabatic stretching energy curve $E_{n_+ n_-}^\text{str} (\theta)$ labeled by the quantum numbers $n_+$ and $n_-$. The angular dependence of the lowest curve $E_{00}(\theta)$ is not resolved at this scale.}
\label{fig:spectrum}
\end{figure}

\begin{figure}[h]
\includegraphics[width=\imagewidth]{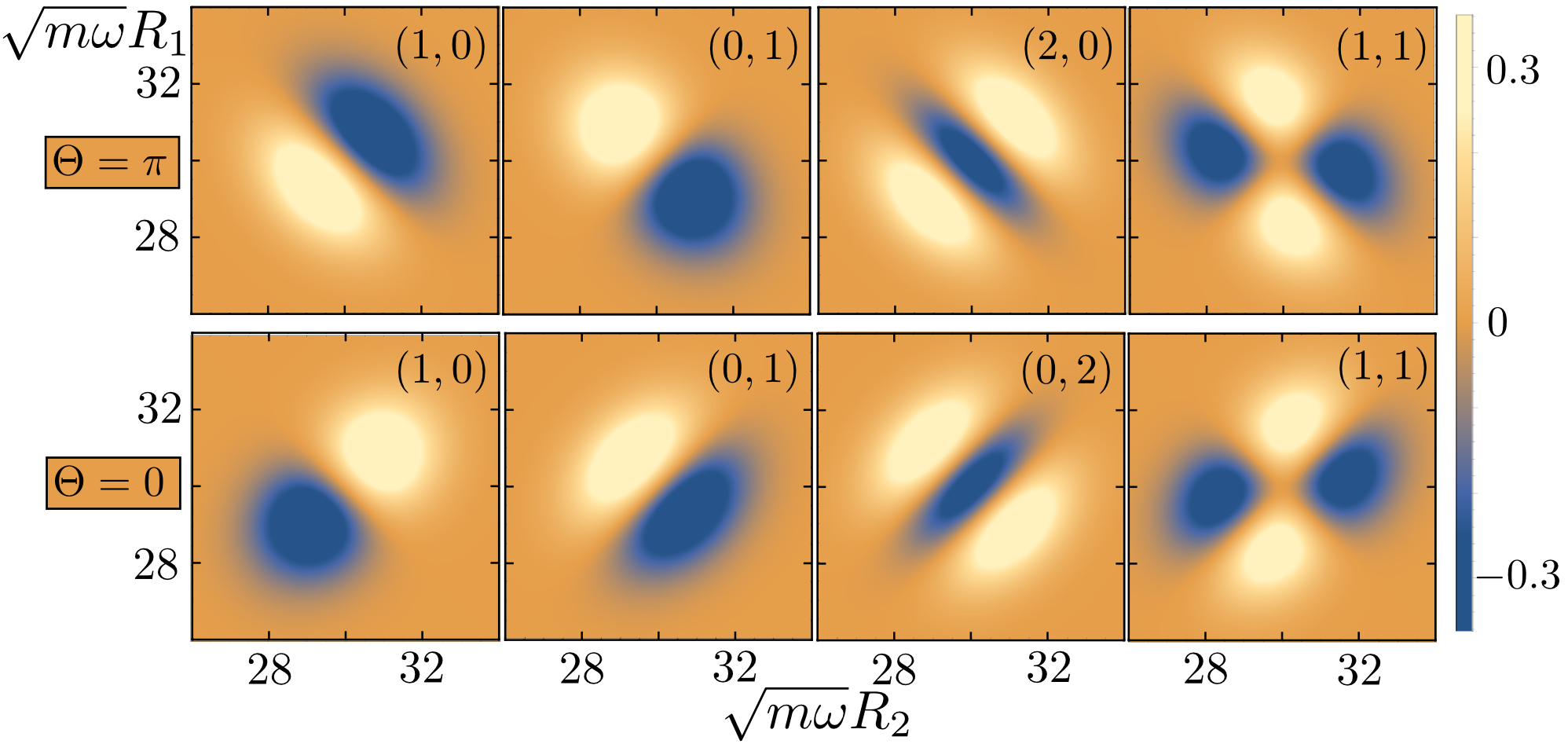}
\caption{Wave functions $\chi^\text{str}_{n_+ n_-}(R_1,R_2;\theta)$  of the stretching modes labeled by the quantum numbers ($n_+$,$n_-$) at the fixed angles $\theta=0$ and $\theta=\pi$. The bond length parameter is set to $\sqrt{m \omega}l_0=30$.} 
\label{fig:stretchmodes}
\end{figure}

To exemplify we consider a case where the bond length parameter is fixed to $\sqrt{m \omega}l_0=30$ which is a realistic ratio for the outer potential wells in the PES of the ULRM analyzed in this work. 
The energy curves in equation (\ref{eqn:energy_toy}) of the lowest stretching modes as well as the corresponding wave functions are presented in Fig.\ \ref{fig:00}, Fig.\ \ref{fig:spectrum} and Fig.\ \ref{fig:stretchmodes}.
%%%%%%%%%%%%%%%%%%%%%%%%%%%%%%%%%%%%%%%%%%%%%%%%%%%%%%%
%%%%%%%%%%%%%%%%%%%%%%%%%%%%%%%%%%%%%%%%%%%%%%%%%%%%%
%%%%%%%%%%%%%%%%%%%%%%%%%%%%%%%%%%%%%%%%%%%%%%%%%%%%%%%%
%%%%%%%%%%%%%%%%%%%%%%%%%%%%%%%%%%%%%%%%%%%%%%%%%%%%%%%%%%
The curve $E_{00}^\text{str}(\theta)$ in Fig.\ \ref{fig:00} possesses two minima at the positions $\theta=0$ and $\theta=\pi$, where the kinetic coupling term in (\ref{eqn:Hamiltonian_strechting}) is the strongest, and a maximum at the position $\theta=\pi/2$ where the kinetic coupling vanishes. This is a general feature of energy curves with symmetric numbers $n_+=n_-$, whereas asymmetric combinations result in asymmetric curves, visible in Fig.\ \ref{fig:spectrum}. For instance, the minimum of a curve with $n_+>n_-$ lies always at $\theta=\pi$, where the effective mass $m_+$ is maximal and reduces quantum fluctuations in the $R_+$ direction. The general impact on the fluctuations due to the masses  $m_+$ and $m_-$ is visible in Fig.\ \ref{fig:00} and Fig.\ \ref{fig:stretchmodes} as elongation or compression of the wave functions along the diagonal directions $R_+$ and $R_-$.

\begin{figure}[h]
\includegraphics[width=\imagewidth]{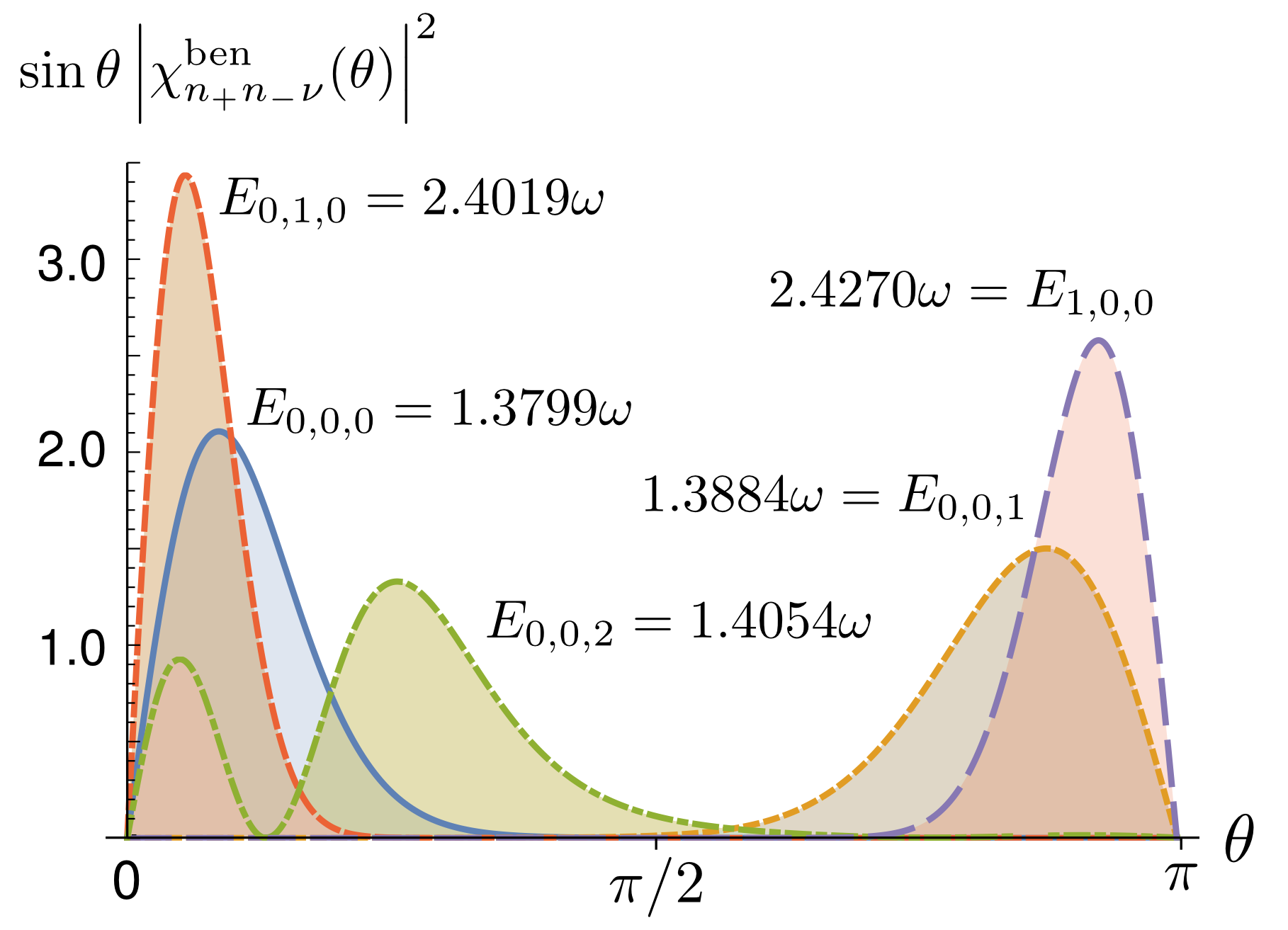}
\caption{Density of the bending modes  $\sin \theta |\chi^\text{ben}_{n_+ n_- \nu}(\theta)|^2$ for the parameter $\sqrt{m \omega}l_0=30$ labeled by their energies $E_{n_+ n_- \nu}$.} 
\label{fig:benmodes}
\end{figure}

The system is expected to localize around angular configurations corresponding to minima in the curves $E_{n_+ n_+}^\text{str}(\theta)$.
To analyze this effect we calculate eigenstates  $\chi^\text{ben}_{n_+ n_-\nu}(\theta)$ of the bending Hamiltonian (\ref{eqn:bending_mode}) for selected curves $E_{n_+ n_+}^\text{str}(\theta)$, where the quantum number $\nu$ labels the excitations in the $\theta$ direction.
Technically this is done by diagonalizing the bending Hamiltonian in a finite basis set consisting of typically 100 Legendre Polynomials $\sqrt{(2l+1)/2} P_l(\cos\theta)$ satisfying the required boundary conditions $\frac{\partial}{\partial \theta} P_l(\cos \theta)=0$ at $\theta=0$ and $\theta=\pi$ \cite{handy_variational_1982}.
The resulting angular densities and energies for some of the energetically lowest eigenstates are depicted in Fig.\ \ref{fig:benmodes}. For instance, for $n_+=n_-$, the two lowest states ($\nu=0$ and $\nu=1$) localize around $\theta=0$ and $\theta=\pi$ and have energies of approximately $1.380 \, \omega$ and $1.389 \, \omega$. 
%%%%%%%%%%%%%%%%%%%%%%%%%%%%%%%%%%%%%%%%%
%%%%%%%%%%%%%%%%%%%%%%%%%%%%%%%%%%%%%%%%%%%%%%%%%%
%
Their difference in energy as well as their difference in the width of the wave functions results from the $\theta$-dependence of the operator $I_j$ in equation (\ref{eqn:moment_of_inertia}).
For the higher states with $n_+=1,n_-=0$ and $n_+=0,n_-=1$, similar effects can be observed. Furthermore the wave functions localize stronger due to the more pronounced minima in the potential curves $E_{n_+ n_+}^\text{str}(\theta)$.
By performing additional MCTDH calculations we verified that the densities of the here discussed vibrational states are in very good agreement with the exact MCTDH results and that the obtained energies differ by not more than a few hundredth of percent. Additionally we validated that an even better accuracy between both methods is achieved when going to higher $\sqrt{m\omega} l_0$, which agrees with our discussion of the adiabatic approximation in appendix \ref{sec:adiabatic}.  
A particular result of this analysis is that, although the interaction potential  $\epsilon(R_1,R_2)$ is isotropic, the energetically lowest eigenstates $\chi^\text{str}_{00}(R_1,R_2;\theta) \chi^\text{ben}_{000}(\theta)$ and $\chi^\text{str}_{00}(R_1,R_2;\theta) \chi^\text{ben}_{001}(\theta)$ are due to the kinetic coupling present in (\ref{eqn:Hamiltonian_strechting}) non-isotropic and localize around $\theta=0$ or $\theta=\pi$.

Some physical intuition concerning the angular configuration of the bending states can be obtained by comparing the quantum system to its classical analogue: three particles connected by two springs with equilibrium length $l_0$ and spring constant $k=\omega^2 m$, cf. equation (\ref{eqn:model_pot}). Although the system possesses equilibrium positions with $R_1^\text{eq}=R_2^\text{eq}=l_0$ at arbitrary angles $\theta^\text{eq}$, only configurations with $\theta^\text{eq}_-=0$ and $\theta^\text{eq}_+=\pi$ are stable against small radial displacements. Contrarily, for all other equilibrium configurations radial displacements will induce not only stretching but also bending oscillations. The frequencies of these bending oscillations can be obtained by a classical adiabatic analysis separating the stretching dynamics with frequencies $\omega_\pm(\theta)=\omega\sqrt{2\pm\cos \theta}$ from the comparably slower bending dynamics. In particular one can show that small oscillations in the symmetric stretching mode with an initial amplitude $\delta R^0_+$ close to the configuration $\theta^\text{eq}_+$ induce bending oscillations around $\theta^\text{eq}_+$ with frequency $\Omega^\text{ben}_+=\sqrt{3/2} \, \omega \delta R^0_+/l_0$ whereas small oscillations in the antisymmetric stretching mode with an initial amplitude $\delta R^0_-$ close to the configuration $\theta^\text{eq}_-$ induce bending oscillations around $\theta^\text{eq}_-$  with frequency $\Omega^\text{ben}_-=\sqrt{1/2} \, \omega \delta R^0_-/l_0$.
The scaling of the bending frequencies $\propto \omega \delta R^0_\pm/l_0$ points out that the bending motion is slow compared to the stretching motion if $\delta R^0_\pm/l_0 \ll1$ and the fact that $\Omega^\text{ben}_+ \neq \Omega^\text{ben}_- $ agrees well with the observation that the energies of the states $\chi^\text{str}_{00}(R_1,R_2;\theta) \chi^\text{ben}_{000}(\theta)$ and $\chi^\text{str}_{00}(R_1,R_2;\theta) \chi^\text{ben}_{001}(\theta)$ are non-degenerate. 

\subsection{ULRM vibrational states}
\label{sec:nuclear_dynamics_ulrm}
In the following we apply the adiabatic approach developed in section \ref{sec:nuc_dynamics_theoret_approach} to determine the stretching and bending motion for selected species of triatomic ULRM. As numerical methods we employ as a first step a 2D finite difference scheme with hard wall boundary conditions to diagonalize the stretching Hamiltonian (\ref{eqn:Hamiltonian_strechting}) for typically 100 different fixed interatomic angles $\theta$ between $0$ and $\pi$. This permits us to determine the stretching wave functions $\chi_j^\text{str}(R_1,R_2;\theta)$, their energy curves $E_j^\text{str}(\theta)$ as well as their other $\theta$-dependent expectation values apparent in equation (\ref{eqn:bending_mode}), like e.g. $\left<1/R_1 R_2\right>_j$.
In contrast to our analysis in section \ref{sec:model} we consider here only wave functions $\chi_j^\text{str}(R_1,R_2;\theta)$ having bosonic symmetry as we investigate ULRM build up by atoms having integer total spin. As a second step we diagonalize the bending Hamiltonian (\ref{eqn:bending_mode}) in a basis set consisting of 100 Legendre Polynomials $\sqrt{(2l+1)/2} P_l(\cos\theta)$. This is done for selected stretching states $\chi_j^\text{str}(R_1,R_2;\theta)$ and yields their bending wave functions $\chi_{j\nu}^\text{ben}(\theta)$ as well as the total energies $E_\nu$ of the molecular states.
With our adiabatic approach we focus exclusively on energetically low-lying vibrational states localized in configurations where $E_j^\text{str}(\theta)$ does not vary too strongly,  although the numerical diagonalization of (\ref{eqn:bending_mode}) yields more states. Especially we do not investigate states with $\vec{R_1} \approx \vec{R_2}$ (see section \ref{sec:PES_trimers}). The accuracy of the adiabatic approximation is verified by comparing the resulting vibrational states to the exact solutions obtained via the MCTDH method.

\subsubsection{Electronic $s$-states of Rubidium}     
\begin{figure}[h]
\includegraphics[width=\imagewidth]{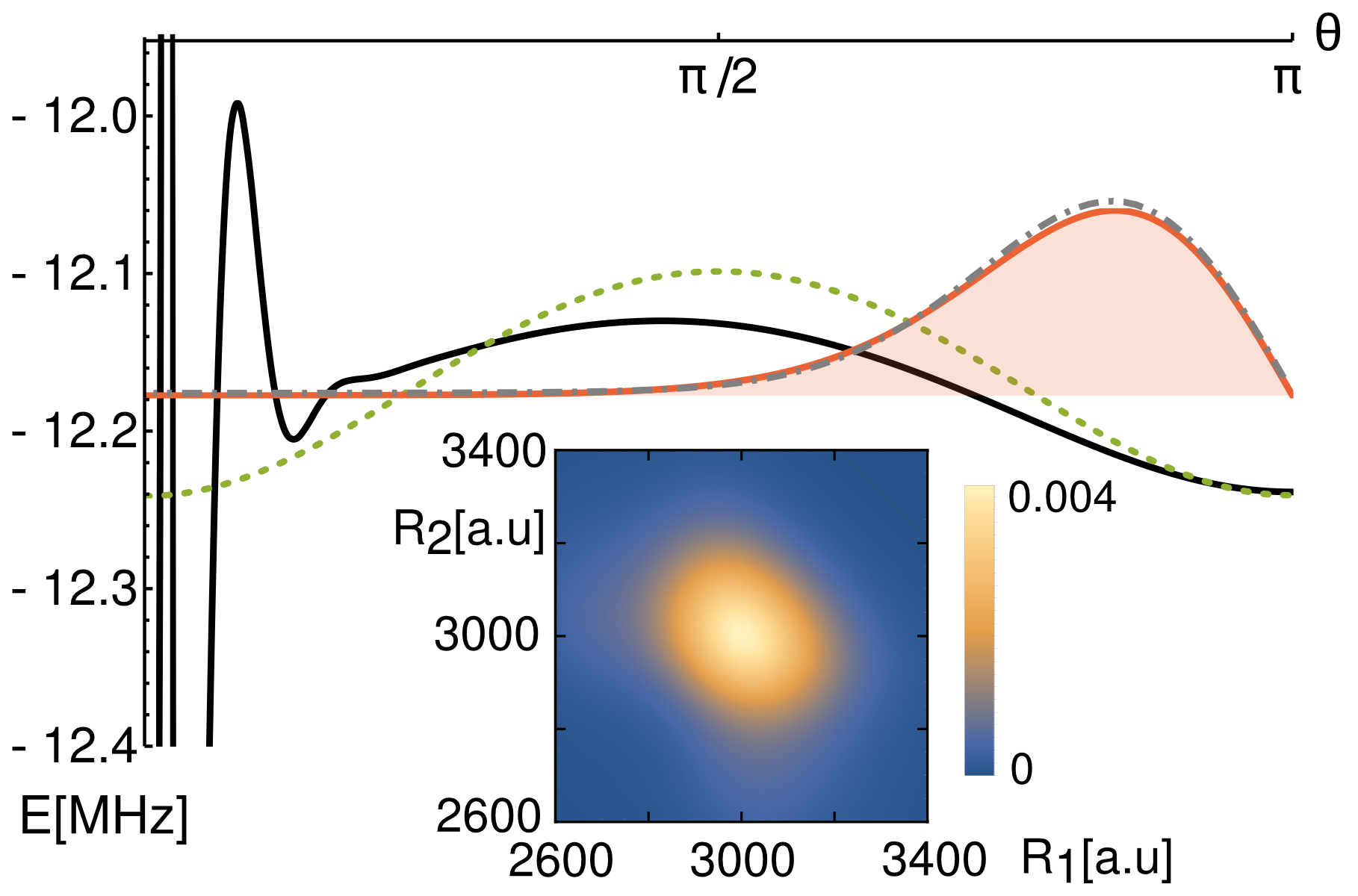}
\caption{Stretching and bending states for the triatomic Rb $43s$ ULRM. The lowest stretching potential energy curve $E_0^\text{str}(\theta)$ (black continuous line) supports the bending wave function with density $\sin \theta |\chi_{00}^\text{ben}(\theta)|^2$ (red filled curve) which is compared to the averaged angular density of the MCTDH solution (gray dashed-dotted line). The inset depicts an image of the stretching state $\chi_0^\text{str}(R_1,R_2;\theta)$ at $\theta=\pi$. Employing the first order perturbation theory PES with effective scattering parameters $a_s=-16.26 \text{ a.u.}$ and $a_p=-25 \text{ a.u.}$ yields an approximate stretching energy curve (green dashed line).
} 
\label{fig:benmodes_rb_40}
\end{figure}

Firstly we consider the triatomic Rb $43s$ state whose PES is presented in Fig.\ \ref{fig:PES_rb40s} and Fig.\ \ref{fig:PES_rb40_rt}.
The properties of the corresponding diatomic system can be deduced from our PES in the limit $R_2 \to \infty$. Qualitatively similar to the Rb $35s$ ULRM discussed in \cite{bendkowsky_rydberg_2010}, the diatomic PES supports one vibrational state localized in the outer well at approximately $R_1 \approx 3000$ a.u.\ with an energy of $-6.06$ MHz as well as several localized and delocalized resonances bound by quantum reflection at the steep potential drop due to the $p$-wave shape resonance. 
Similarly, the spectrum of the stretching Hamiltonian of the triatomic system includes one solution $\chi_0^\text{str}(R_1,R_2;\theta)$ describing a configuration where the two ground state atoms are localized in the outer potential well at $R_1\approx R_2 \approx 3000 \text{ a.u.}$ as well as several resonances bound by quantum reflection. As it is computationally involved to describe these resonances within our finite difference scheme we focus here on the solution $\chi_0^\text{str}(R_1,R_2;\theta)$ whose stretching potential energy curve $E_{0}^\text{str}(\theta)$ and wave function $\chi_0^\text{str}(R_1,R_2;\pi)$ are depicted in Fig.\ \ref{fig:benmodes_rb_40}.
The $\theta$-dependence of the energy curve $E_{0}^\text{str}(\theta)$ results from the interplay of two reasons: 
the kinetic coupling in the stretching Hamiltonian (\ref{eqn:Hamiltonian_strechting}) and the angular dependence of the PES $\epsilon(R_1,R_2,\theta)$. As discussed in section \ref{sec:PES_trimers} the latter dominates at angles close to zero and modifies $E_{0}^\text{str}(\theta)$ strongly around $\theta=0$. At larger angles $\epsilon(R_1,R_2,\theta)$ depends much weaker on $\theta$ and the influence of the kinetic coupling becomes important. 
In this region the stretching energy lies roughly around twice the dimer binding energy (6.06 MHz) and $E_0^\text{str}(\theta)$ as well as $\chi_0^\text{str}(R_1,R_2;\theta)$ resemble in their appearance the results of the model Hamiltonian in Fig.\ \ref{fig:00}. In particular $E_0^\text{str}(\theta)$ has a minimum at $\theta=\pi$ and $\chi_0^\text{str}(R_1,R_2;\theta=\pi)$ is elongated along the $R_1=-R_2$ diagonal.

To quantify the impact of the residual anisotropy of $\epsilon(R_1,R_2,\theta)$ on $E_0^\text{str}(\theta)$ in the region of larger $\theta$ we compare $E_0^\text{str}(\theta)$ in Fig.\ \ref{fig:benmodes_rb_40} to the energy curve obtained by using the isotropic PES from the first order result in equation (\ref{eqn:energy_first_order_sp}) with effective non-energy-dependent scattering parameters $a_s=-16.26 \text{ a.u.}$ and $a_p=-25 \text{ a.u.}$ which are chosen to fit the exact PES at $\theta=\pi$. This curve describes qualitatively well the shape of $E_0^\text{str}(\theta)$ for $\theta > \pi/4$ which points out that the anisotropy of $\epsilon(R_1,R_2,\theta)$ is indeed weak and that perturbation theory is a useful approximation in this part of configuration space. 

With our adiabatic bending state analysis we focus on states localized around the minimum of $E_0^\text{str}(\theta)$ at $\theta=\pi$. The corresponding energetically lowest state $\chi_{00}^\text{ben}(\theta)$ with an energy of approximately $-12.18$ MHz is depicted in Fig.\ \ref{fig:benmodes_rb_40}. The ULRM in this state are close to a linear structure and the probability to detect angles $\theta < \pi/2$ is nearly zero.   
The energetically next higher bending state with $-12.14$ MHz obtained in our diagonalization procedure leaks into the region around $\theta=0$ and is not displayed.
In order to compare the adiabatic solution to the exact solution of the vibrational Hamiltonian $\ref{eqn:Hamiltonian_nuclear}$, we depict in Fig.\ \ref{fig:benmodes_rb_40} also the energy and the angular density of the MCTDH solutions, i.e. the density of the nuclear wave function averaged over $R_1$ and $R_2$. For the discussed state both methods are in excellent agreement.

These results firstly confirm the observation that there are triatomic $l=0$ ULRM having to a high accuracy twice the energy of the dimer states \cite{bendkowsky_rydberg_2010,gaj_molecular_2014} and secondly, it is legitimate to assign a certain geometric structure to these states. 
Furthermore, in the context of the recent work \cite{schlagmuller_probing_2016}, the stretching function $E_0^\text{str}(\theta)$ could be applied to gain information on the profile of trimer peaks in experimental spectra by performing corresponding classical calculations.
  
\subsubsection{Electronic $s$-states of Strontium}    
As a second example we discuss the bending and stretching dynamics of a triatomic ULRM consisting of a $^{84}$Sr atom in an electronic triplet $5s33s$ Rydberg state interacting with two $^{84}$Sr ground state atoms via $s$- and $p$-wave scattering.
Due to the absence of the $p$-wave shape resonance the PES of the Sr system supports more bound stretching solutions than the corresponding Rb molecule and possesses therefore a richer variety of vibrational states. The diatomic system has been analyzed in \cite{desalvo_ultra-long-range_2015} within a two-active-electron model.
It has been shown that experimental spectra can be reproduced within the Fermi pseudopotential approach by employing a $\Delta_0=3.376$ quantum defect single electron Coulomb wave function and effective scattering lengths $a_s[k]=a_s[0] + \frac{\pi}{3} \alpha \,k$ and $a_p[k]=a_p[0]$, where $a_s[0]=-13.2 \text{ a.u.}$, $\alpha=186 \text{ a.u.}$ and $a_p[0]= -25 \text{ a.u.}$.
The energies of the three lowest diatomic vibrational states $\chi^\text{dim}_0$, $\chi^\text{dim}_1$ and $\chi^\text{dim}_2$ were determined to -25.0 MHz, -11.1 MHz and -8.6 MHz. 
In contrast to the states $\chi^\text{dim}_0$ and $\chi^\text{dim}_1$ being localized in the outer well at $1650 \text{ a.u.}$, the state $\chi^\text{dim}_2$ is delocalized over the three outer potentials wells at approximately $1650\text{ a.u.}$, $1400\text{ a.u.}$ and $1250 \text{ a.u.}$.       
 Here we adapt these results to calculate the triatomic PES approximately via equation (\ref{eqn:energy_first_order_sp}) which is simply the sum of the diatomic potentials used in \cite{desalvo_ultra-long-range_2015}.  
According to our previous analysis this procedure should describe well the vibrational structure of states localized sufficiently far from the $\theta=0$ configuration.

%%%%%%%%%%%%%%%%%
\begin{figure}[h]
\includegraphics[width=\imagewidth]{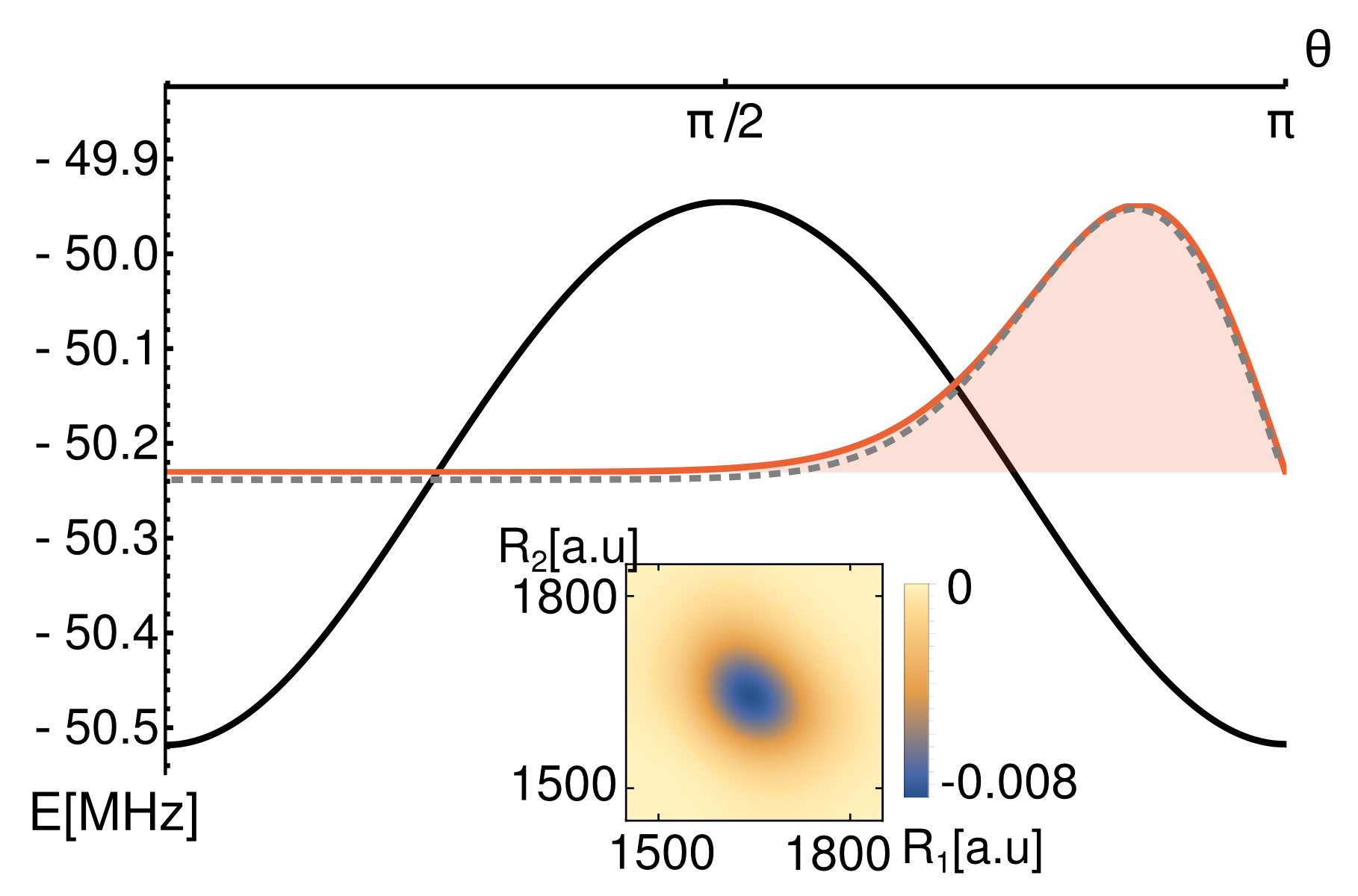}
\caption{Stretching and bending state for a triatomic $^{84}$Sr  ULRM in a $n=33$ Rydberg state. The lowest stretching curve $E_0^\text{str}(\theta)$ (black line) supports the bending state with density $\sin \theta |\chi_{00}^\text{ben}(\theta)|^2$ (red filled curve) which is compared to the averaged angular density of the MCTDH solution (gray dashed line). The inset depicts the stretching state $\chi_0^\text{str}(R_1,R_2;\theta)$ at $\theta=\pi$.}
\label{fig:30sr_ben_0} 
\end{figure}
%%%%%%%%%%%%%%%%%%%%%
\begin{figure}[h]
\includegraphics[width=\imagewidth]{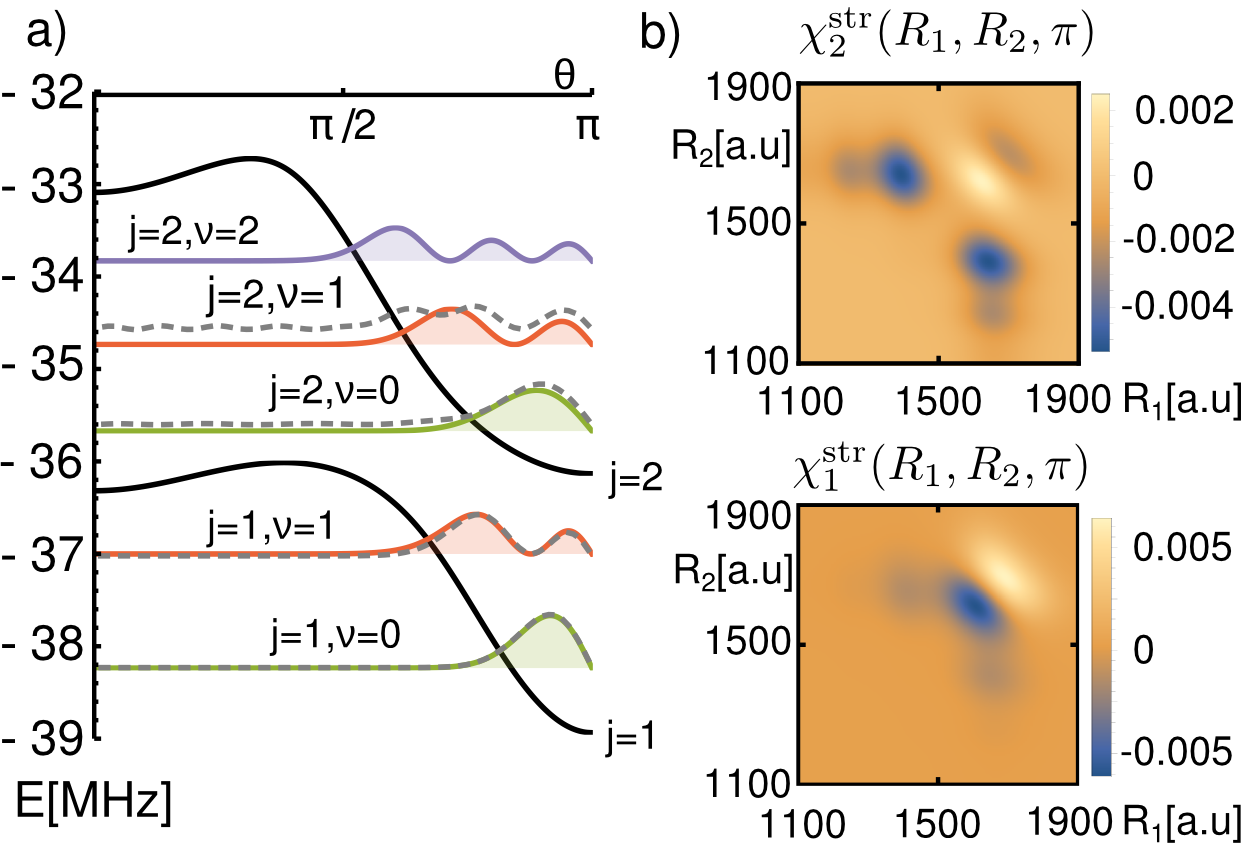}
\caption{a) Energy curves $E^\text{str}_j(\theta)$ of the $j=1$ and $j=2$ stretching states (black lines) for a triatomic $^{84}$Sr ULRM in a $n=33$ Rydberg state. Each curve supports bending states $\chi_{j\nu}$ represented by their angular densities $\sin \theta |\chi_{j\nu}^\text{ben}(\theta)|^2$ (colored filled curves) which are compared to the densities of the MCTDH solutions (gray dashed line). b) The insets depict the wave functions of the stretching states $\chi_1^\text{str}$ and $\chi_2^\text{str}$ at $\theta=\pi$. } 
\label{fig:30sr_ben_12}
\end{figure}
%%%%%%%%%%%%%%%%%%%%%%%%%

The energetically lowest stretching states are depicted in Fig.\ \ref{fig:30sr_ben_0} and Fig.\ \ref{fig:30sr_ben_12}.
Their energy curves can be interpreted as follows: At the angle $\theta=\pi/2$ the kinetic coupling present in (\ref{eqn:Hamiltonian_strechting}) vanishes and $H^\text{str}$ becomes separable. Consequently, all stretching states $\chi^\text{str}_j(R_1,R_2;\pi/2)$ can be written as bosonic wave functions $\chi^\text{str}_j(R_1,R_2;\pi/2)=\frac{1}{\sqrt{2}}\left(\chi^\text{dim}_{\nu_1}(R_1)\chi^\text{dim}_{\nu_2}(R_2) +\chi^\text{dim}_{\nu_1}(R_2)\chi^\text{dim}_{\nu_2}(R_1) \right)$ build up by product states of two diatomic vibrational states $\chi^\text{dim}_{\nu_1}$ and $\chi^\text{dim}_{\nu_2}$.  
All curves $E^\text{str}_j(\theta)$ evaluated at $\theta=\pi/2$ yield therefore the sum of the energies of two diatomic vibrational states.
At angles $\theta\neq \pi/2$ the kinetic coupling mixes these states and deforms the energy curves, i.e. the curves $E^\text{str}_j(\theta)$ are not constant. 
However, the dominant underlying diatomic vibrational modes $\chi^\text{dim}_{\nu_1}$ and $\chi^\text{dim}_{\nu_2}$ can still be identified and characterize the stretching state. This interpretation is similar to the shell model introduced in \cite{schmidt_mesoscopic_2016}.
For example the state $\chi^\text{str}_0(R_1,R_2,\pi)$ in Fig.\ \ref{fig:30sr_ben_0} has roughly twice the binding energy of the lowest diatomic state and describes a situation where the two ground state atoms are bound in the lowest vibrational modes $\chi^\text{dim}_0$ at distances $R_1\approx R_2 \approx 1650$ a.u..  
This stretching mode supports one bending state around $\theta=\pi$ and resembles in its characteristics very much the Rb state in Fig.\ \ref{fig:benmodes_rb_40}.
The next higher stretching state $\chi^\text{str}_1(R_1,R_2,\pi)$ can be viewed as a combination of the first two diatomic vibrational modes $\chi^\text{dim}_0$ and $\chi^\text{dim}_1$. It resembles the state $\chi^\text{str}_{10}(R_1,R_2,\pi)$ of the model system depicted in Fig.\ \ref{fig:stretchmodes} which is excited in the symmetric stretching mode. In a close analogy the stretching  potential energy $E^\text{str}_1(\theta)$ possesses a minimum at $\theta=\pi$. 
Finally, the stretching state $\chi^\text{str}_2$ is approximately a combination of the diatomic vibrational modes $\chi^\text{dim}_0$ and $\chi^\text{dim}_2$. It possesses also a distinct energy minimum at $\theta=\pi$ supporting several bending states. Close to this minimum it describes a delocalized state with high probability to find one atom situated in the outer well at $R_1\approx 1650$ a.u.\ and the other one at $R_2\approx 1400$ a.u..

Again, the adiabatic nuclear wave functions can be compared to the energies and angular densities of the MCTDH solutions which are depicted in Fig.\ \ref{fig:30sr_ben_0} and Fig.\ \ref{fig:30sr_ben_12}. While there is excellent agreement for all presented bending states $\chi^\text{ben}_{j\nu}$ with $j=0$ and $j=1$,  it becomes evident that bending states with $j=2$ do not have accurate counterparts in the MCTDH results, e.g. we could not identify any state similar to $\chi^\text{str}_{2}(R_1,R_2;\theta) \chi^\text{ben}_{22}(\theta)$. 
This indicates that there is a breakdown of the adiabatic approximation for these states which is also signaled by the avoided crossing behavior of the stretching curves $E^\text{str}_1(\theta)$ and $E^\text{str}_2(\theta)$ visible in Fig.\ \ref{fig:30sr_ben_12}. However, the adiabatic approximation works well for the lowest bending states in the stretching mode  $\chi^\text{str}_1$. Their energetic spacing is on the order of $\sim 1$ MHz which is comparable to the $\sim 800$ KHz linewidth of the laser \cite{desalvo_ultra-long-range_2015} and might be resolvable in future experiments.

\subsubsection{Electronic $p$-states of Rubidium}
To outline qualitative changes when going to higher $l$ states we discuss lastly the vibrational structure of the triatomic Rb $42p$ state with the two PES $\epsilon_-$ and $\epsilon_+$  shown in Fig.\ \ref{fig:PES_rb40p} and Fig.\ \ref{fig:PES_rb40p_theta}. We focus here on vibrational states bound in the outer potential well of the lower PES $\epsilon_-(R_1,R_2,\theta)$ at $R_1=R_2\approx 2930 \text{ a.u.}$ and $\theta=\pi$. 
%.
In the limit $R_2\to \infty$ we obtain the two lowest vibrational states $\chi^\text{dim}_0$ and $\chi^\text{dim}_1$ of the corresponding diatomic molecule having vibrational energies of -18.00 MHz and -8.70 MHz. The state $\chi^\text{dim}_0$ is localized in the outer potential well at $2930 \text{ a.u.}$ whereas  $\chi^\text{dim}_1$ is located in the outer well but also to a small fraction in the second outer well around $2550 \text{ a.u.}$.  

\begin{figure}[h]
\includegraphics[width=1\imagewidth]{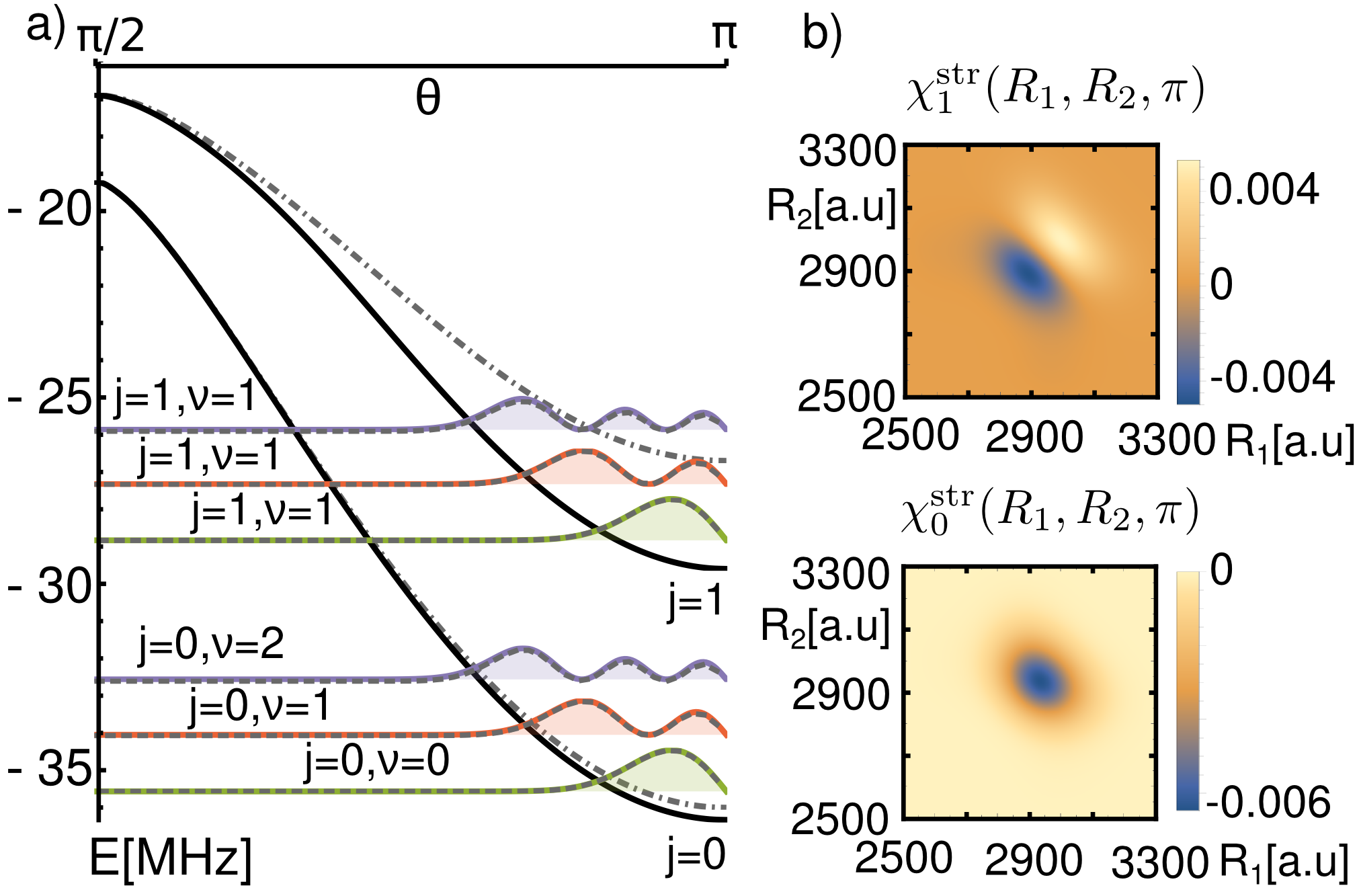}
\caption{a) Energy curves $E^\text{str}_j(\theta)$ (continuous black lines) of the $j=0$ and $j=1$ stretching states for the triatomic $42p$ ULRM in comparison to the curves $\tilde{E}^\text{str}_j(\theta)$ (dashed dotted gray lines) neglecting the kinetic coupling. Each curve $E^\text{str}_j(\theta)$ supports bending states $\chi_{j\nu}^\text{ben}$ represented by their angular densities $\sin\theta |\chi_{j\nu}^\text{ben}(\theta)|^2$ (colored filled curves) which are compared to the MCTDH solutions (gray dashed lines)  b) The insets depict the wave functions of the stretching states $\chi_0^\text{str}$ and $\chi_1^\text{str}$ at $\theta=\pi$}
\label{fig:rb40p_ben} 
\end{figure}

The two energetically lowest stretching states of the triatomic ULRM and their energy curves are depicted in Fig.\ \ref{fig:rb40p_ben} for angles $\theta \geq \pi/2$.
Both energy curves possess a minimum at $\theta=\pi$ where they support several bending states.
The stretching state $\chi_0^\text{str}$ around $\theta=\pi$ describes approximately a situation where the two ground state atoms are bound in the lowest diatomic vibrational mode $\chi^\text{dim}_0$ whereas $\chi_1^\text{str}$ corresponds approximately to a situation where one atom is bound in the $\chi^\text{dim}_0$ mode while the second atom is bound in the $\chi^\text{dim}_1$ mode. This characterization holds only approximately as the stretching Hamiltonian (\ref{eqn:Hamiltonian_strechting}) is not separable with respect to $R_1$ and $R_2$.  

In contrast to the $l=0$ ULRM the minimum of the curves $E^\text{str}_j(\theta)$ at $\theta=\pi$ is in this case caused mainly by the non-isotropic PES. To quantify the impact of the kinetic coupling we compare $E^\text{str}_j(\theta)$ to the curves $\tilde{E}^\text{str}_j(\theta)$ which are obtained by setting the kinetic coupling operator in $H^\text{str}$ artificially to zero.
Under this constraint all curves $\tilde{E}^\text{str}_j(\theta)$ evaluated at $\theta=\pi$ equal the sum of two diatomic vibrational energies which needs to be the case as the PES $\epsilon_-$ equals at $\theta=\pi$ the sum of the diatomic PES (see section \ref{sec:PES_trimers}).
The comparison to the full stretching potential energy curves $E^\text{str}_j(\theta)$ demonstrates that the kinetic coupling operator lowers the potential wells more importantly for the $\chi^\text{str}_1$ state than for the $\chi^\text{str}_0$ state. This can be understood from our analysis of the model system which points out that energy curves of stretching states having more nodes along the $R_1=R_2$ diagonal than along the $R_1=-R_2$ off-diagonal are lowered stronger by the kinetic coupling at $\theta=\pi$ than the energy curves of stretching states having no nodes.

These results examine some of the basic characteristics of triatomic $l=1$ ULRM.  However, it is expected that additional corrections due to the spin dependent scattering channels, the fine structure of the Rydberg atom and the hyperfine structure of the ground state atoms will modify the electronic and vibrational structure. Studying these effects for polyatomic $l=1$ and $l=2$ ULRM is an interesting subject for future investigations.   

\section{Conclusion}
We proposed an approach to solve the electronic problem for polyatomic ULRM with $N$ ground state atoms bound by $s$- and $p$-wave contact interaction. The method is equivalent to the finite basis set diagonalization but reduces numerical efforts by employing the Green's function of the unperturbed Rydberg system.
This method was applied to determine the potential energy surfaces of triatomic Rb ULRM in electronic $l=0$ and $l=1$ states.
In particular we analyzed the impact of high-$l$ admixtures on the PES.
We found that, from a perturbative point of view, higher order corrections can disturb the first order PES importantly when the ground state atoms are close together but do not induce qualitative changes in other configurations.

Approximate solutions to the vibrational problem of the triatomic systems determining not only the radial but also the angular configuration of the nuclei were determined by separating the bending and stretching motion adiabatically. In our analysis of these states we focused on the impact of kinetic couplings apparent in the vibrational Hamiltonian and showed that these terms energetically favour linear configurations of the nuclei.
We quantified this effect for triatomic Rb and Sr ULRM in $l=0$ states as well as for Rb ULRM in $l=1$ states.
The coupling influences the molecular geometry most importantly if the PES depends only weakly on the interatomic angle, e.g. for $l=0$ states, and if the two ground state atoms are bound in different diatomic vibrational modes. These findings specify the geometry of experimentally observed triatomic $l=0$ ULRM. By performing additional numerically exact MCTDH calculations we verified the accurateness of the adiabatic analysis. The next step towards an ameliorated understanding of this system would be to improve the electronic Hamiltonian by including additional interactions like angular momentum couplings and interactions between ground state atoms.

The universal binding mechanism in ULRM allows principally also the formation of larger polymers including several ground state atoms.
Like their diatomic counterparts, these systems are expected to be extremely sensitive to electric and magnetic fields and open therefore unique possibilities to control molecular properties like the geometry, the orientation or the electric dipole moment via weak electric and magnetic fields. The proposed Green's function approach should be very suitable for studying such effects in future investigations.

%%%%%%%%%%%%%%%%%%%%%%%%%%%%%%%%%%%%%%%%%%%%%%%%%%%%%%%%%%%%%%%%%%%%%%%%%%%%%
\begin{acknowledgments}
We thank S. T. Rittenhouse for fruitful discussions concerning the Green’s function approach and we thank H.-D. Meyer
for providing us very helpful supplementary material on the
MCTDH package.
C.F. gratefully acknowledges a scholarship by the Studienstiftung des deutschen Volkes.
\end{acknowledgments}

%%%%%%%%%%%%%%%%%%%%%%%%%%%%%%%%%%%%%%%%%%%%%%%%%%%%%%%%%%%%%%%%%%%%%%%%%%%%%

%\bibliographystyle{unsrt}
\bibliographystyle{apsrev4-1}
\bibliography{final}
%\begin{thebibliography}{}
  % \input{final.bbl}
%\end{thebibliography}

%%%%%%%%%%%%%%%%%%%%%%%%%%%%%%%%%%%%%%%%%%%%%%%%%%%%%%%%%%%%%%%%%%%%%%%%%%%%%
\appendix
\section{Justification of the adiabatic approximation}
\label{sec:adiabatic}
Here we derive the effective bending Hamiltonian (\ref{eqn:bending_mode}) from the vibrational Hamiltonian $H^\text{vib}$ in (\ref{eqn:Hamiltonian_nuclear}). 
Formally similar to the Born-Huang approach to the Born-Oppenheimer separation \cite{drake_molecular_2006} the Schr\"odinger equation $H^\text{vib} \chi_\nu (R_1,R_2,\theta) =E_\nu \chi_\nu (R_1,R_2,\theta)$ with energy $E_\nu$ can be cast into a system of coupled channel equations by inserting the expansion (\ref{eqn:adiabatic_expansion}) and projecting it onto the eigenstates  $\chi^\text{str}_i(R_1,R_2;\theta)$ of $H^\text{str}$ with eigenvalues $E_i^\text{str}(\theta)$. This system has the form 

\begin{equation}
0=\sum_j \left[A_{ij}+B_{ij} + \delta_{ij} \left(E_j^\text{str}(\theta)-E_\nu\right) \right] \chi_{j\nu}^\text{ben}(\theta)  \ ,
\label{eqn:bend_appendix}
\end{equation}
where the operators $A_{ij}$ and $B_{ij}$ are given by

\begin{align}
A_{ij}=& -\frac{1}{m} \left( \bra{\chi^\text{str}_i} \frac{1}{R_1^2}\ket{\chi^\text{str}_j}+ \bra{\chi^\text{str}_i} \frac{1}{R_2^2}\ket{\chi^\text{str}_j}- \bra{\chi^\text{str}_i} \frac{1}{R_1 R_2}\ket{\chi^\text{str}_j} \cos \theta \right)\left(\frac{\partial^2}{\partial \theta^2} + \cot \theta \frac{\partial}{\partial \theta}  \right)  \nonumber \\
&-\frac{1}{m} \left(\bra{\chi^\text{str}_i} \frac{1}{R_1 R_2}\ket{\chi^\text{str}_j}
-\bra{\chi^\text{str}_i}\frac{1}{R_2}\frac{\partial}{\partial R_1}\ket{\chi^\text{str}_j}
-\bra{\chi^\text{str}_i}\frac{1}{R_1}\frac{\partial}{ \partial R_2}\ket{\chi^\text{str}_j}
\right)\left(\cos \theta + \sin \theta \frac{\partial}{\partial \theta} \right)
\end{align}
and

\begin{align}
B_{ij}=& -\frac{1}{m} \left( \bra{\chi^\text{str}_i} \frac{1}{R_1^2}\ket{\partial_\theta \chi^\text{str}_j}+ \bra{\chi^\text{str}_i} \frac{1}{R_2^2}\ket{\partial_\theta \chi^\text{str}_j}- \bra{\chi^\text{str}_i} \frac{1}{R_1 R_2}\ket{\partial_\theta \chi^\text{str}_j} \cos \theta \right)\left(2 \frac{\partial}{\partial \theta} + \cot \theta  \right)  \nonumber \\
&-\frac{1}{m} \left(\bra{\chi^\text{str}_i} \frac{1}{R_1 R_2}\ket{\partial_\theta \chi^\text{str}_j}
-\bra{\chi^\text{str}_i}\frac{1}{R_2}\frac{\partial}{\partial R_1}\ket{\partial_\theta \chi^\text{str}_j}
-\bra{\chi^\text{str}_i}\frac{1}{R_1}\frac{\partial}{ \partial R_2}\ket{\partial_\theta \chi^\text{str}_j}
\right)\sin \theta  \nonumber\\
 & -\frac{1}{m} \left( \bra{\chi^\text{str}_i} \frac{1}{R_1^2}\ket{\partial^2_\theta \chi^\text{str}_j}+ \bra{\chi^\text{str}_i} \frac{1}{R_2^2}\ket{\partial^2_\theta \chi^\text{str}_j}- \bra{\chi^\text{str}_i} \frac{1}{R_1R_2}\ket{\partial^2_\theta \chi^\text{str}_j} \cos \theta \right)
\end{align}
Neglecting the effects of all operators $B_{ij}$, of all off-diagonal operators $A_{ij}$ with $i\neq j$ and recognizing that certain parts of the diagonal operators $A_{ii}$ vanish leads to the bending Hamiltonian (\ref{eqn:bending_mode}) having solutions. In the following we will point out the necessary steps and discuss conditions under which this approximation is valid. We consider stretching states $\chi_j^\text{str}(R_1,R_2;\theta)$ similar to the solutions of the model Hamiltonian discussed in section \ref{sec:model}. We suppose that these states localize around distances $R_1^\text{eq}$ and $R_2^\text{eq}$ with typical fluctuations in bond length $\Delta r$. Their wave functions $\chi_j^\text{str}(R_1,R_2;\theta)$ are chosen to be real. Furthermore we restrict our analysis to wave functions localized in regions without any crossings of energy curves. For simplicity we will here only discuss the case $R_1^\text{eq}=R_2^\text{eq}:=R^\text{eq}$. 
The crucial assumptions for the adiabatic approximation is that the bond length should be large and satisfy
\begin{equation}
R^\text{eq} \gg \Delta r \ .
\end{equation}
To estimate the order of all non-adiabatic couplings we express the off-diagonal elements in $A_{ij}$ via the identity
\begin{equation}
\bra{\chi^\text{str}_i} f(R_1,R_2) \ket{\chi^\text{str}_j}=\frac{\bra{\chi^\text{str}_i} \left[f(R_1,R_2),H^\text{str}\right]\ket{\chi^\text{str}_j}}{E^\text{str}_j(\theta)-E^\text{str}_i(\theta)}
\label{eqn:aij_offdiagonal}
\end{equation}
where $[\cdot,\cdot]$ denotes the commutator and $f(R_1,R_2)$ needs to be replaced by $1/R_1^2$, $1/R_2^2$, $1/R_1 R_2$, $\partial_{R_1}/R_2$ or $\partial_{R_2}/R_1$. 
The non-adiabatic derivative couplings $B_{ij}$ with $i \neq j$ can be obtained from
\begin{align}
\bra{\chi^\text{str}_i} f(R_1,R_2) \ket{\partial_\theta \chi^\text{str}_j}=&\frac{\bra{\chi^\text{str}_i} \left[f(R_1,R_2)\frac{\partial}{\partial \theta},H^\text{str}\right]\ket{\chi^\text{str}_j}}{E^\text{str}_j(\theta)-E^\text{str}_i(\theta)}
\nonumber \\
&- \frac{(\frac{\partial}{\partial \theta} E^\text{str}_j(\theta))}{E^\text{str}_j(\theta)-E^\text{str}_i(\theta)} \bra{\chi^\text{str}_i} f(R_1,R_2) \ket{\chi^\text{str}_j}
\label{eqn:derivative1_appendix}
\end{align}
and
\begin{equation}
\bra{\chi^\text{str}_i} f(R_1,R_2)\ket{\partial^2_\theta \chi^\text{str}_j}= \sum_{k \neq i} \braket{\chi^\text{str}_i}{\partial_\theta \chi^\text{str}_k} 
\bra{\chi^\text{str}_k} f(R_1,R_2)\ket{\partial_\theta \chi^\text{str}_j} \ ,
\label{eqn:derivative2_appendix}
\end{equation}
where the sum runs in principal over all stretching states as $\braket{\chi^\text{str}_i}{\partial_\theta \chi^\text{str}_i}=0$ for real-valued wave functions. By employing (\ref{eqn:derivative1_appendix}) for $f(R_1,R_2)=1$ one can write (\ref{eqn:derivative2_appendix}) as 
\begin{equation}
\bra{\chi^\text{str}_i} f(R_1,R_2)\ket{\partial^2_\theta \chi^\text{str}_j}= \sum_{k \neq i} \frac{\bra{\chi^\text{str}_i} \left[\frac{\partial}{\partial \theta},H^\text{str}\right]\ket{\chi^\text{str}_k}}{E^\text{str}_k(\theta)-E^\text{str}_i(\theta)}
\bra{\chi^\text{str}_k} f(R_1,R_2)\ket{\partial_\theta \chi^\text{str}_j} \ .
\label{eqn:derivative3_appendix}
\end{equation}
Typically, the energy spacing between adjacent stretching levels can be related to the length scale $\Delta r$ via $E^\text{str}_j(\theta)-E^\text{str}_i(\theta)\approx 1/(m \Delta R^2)$, e.g. this is the case for harmonic-like confinements. The matrix elements containing $f(R_1,R_2)$ scale as $1/R^\text{eq}$ or $(1/R^\text{eq})^2$.  
Therefore all non-adiabatic couplings present in (\ref{eqn:aij_offdiagonal}), (\ref{eqn:derivative1_appendix}) and (\ref{eqn:derivative3_appendix}) scale at least as $\Delta r/R^\text{eq}$. Hence,  in the limit  $\Delta r/R^\text{eq} \to 0$ it is appropriate to consider only the diagonal elements $A_{ii}$ and $B_{ii}$, which is the adiabatic approximation. 

The remaining diagonal operators $A_{ii}$ and $B_{ii}$ can be further simplified. $A_{ii}$ contains elements of the type $\bra{\chi^\text{str}_i}\frac{1}{R_2}\frac{\partial}{\partial R_1}\ket{\chi^\text{str}_i}$. By calculating the adjoint one can show that $\bra{\chi^\text{str}_i}\frac{1}{R_2}\frac{\partial}{\partial R_1}\ket{\chi^\text{str}_i}
=-\bra{\chi^\text{str}_i}\frac{1}{R_2}\frac{\partial}{\partial R_1}\ket{\chi^\text{str}_i}$. Consequently all diagonal elements $\bra{\chi^\text{str}_i}\frac{1}{R_2}\frac{\partial}{\partial R_1}\ket{\chi^\text{str}_i}$ and $\bra{\chi^\text{str}_i}\frac{1}{R_1}\frac{\partial}{\partial R_2}\ket{\chi^\text{str}_i}$  vanish. For the same reason also the terms $\bra{\chi^\text{str}_i}\frac{1}{R_2}\frac{\partial}{\partial R_1}\ket{\partial_\theta \chi^\text{str}_i}$ and $\bra{\chi^\text{str}_i}\frac{1}{R_1}\frac{\partial}{\partial R_2}\ket{\partial_\theta \chi^\text{str}_i}$ in the operator $B_{ii}$ are zero.

Next we will show under which conditions the remaining parts of the diagonal operators $B_{ii}$ are small compared to the corresponding terms in the operators $A_{ii}$ and can be neglected. The operator $B_{ii}$ contains derivatives of the stretching wave function with respect to the angle $\theta$. 
Firstly we focus on terms containing first order derivatives of the form $\left<\chi_i^\text{str}\left|f(R_1,R_2)\right| \partial_\theta \chi^\text{str}_i\right>$ where the observable $f(R_1,R_2)$ needs to be replaced by $1/R_1^2$, $1/R_2^2$ or $1/R_1 R_2$. One can show that 
\begin{equation}
\left<\chi_i^\text{str}\left|f(R_1,R_2)\right| \partial_\theta \chi^\text{str}_i\right>=
\frac{1}{2} \frac{\partial}{\partial \theta} \left<\chi_i^\text{str}\left|f(R_1,R_2)\right|\chi^\text{str}_i\right> 
\end{equation}
and, consequently, these elements of $B_{ii}$ can be neglected (compared to $A_{ii}$) if
\begin{equation}
 \frac{\frac{\partial}{\partial \theta} \left<\chi_i^\text{str}\left|f(R_1,R_2)\right|\chi^\text{str}_i\right>}{\left<\chi_i^\text{str}\left|f(R_1,R_2)\right|\chi^\text{str}_i\right> }\ll 1 \ .
 \end{equation}
This means that the bond lengths should depend only very weakly on the angle $\theta$, which is typically the case in our system. 

Lastly we need to discuss the terms of $B_{ii}$ containing second order derivatives of the form 
$\left<\chi_i^\text{str}\left|f(R_1,R_2)\right| \partial^2_\theta \chi^\text{str}_i\right>$. To this aim we will make use of the fact that $f(R_1,R_2)$ varies slowly in the range of the stretching wave function and approximate $\left<\chi_i^\text{str}\left|f(R_1,R_2)\right| \partial^2_\theta \chi^\text{str}_i\right> \approx f(R^\text{eq},R^\text{eq}) \left<\chi_i^\text{str}|\partial^2_\theta \chi^\text{str}_i\right> $. These terms are small compared to the corresponding terms in $A_{ii}$ if $\left<\chi_i^\text{str}|\partial^2_\theta \chi^\text{str}_i\right> \ll 1$.  By employing (\ref{eqn:derivative1_appendix}) for $f(R_1,R_2)=1$ all terms containing second order derivatives can be obtained from the identity
\begin{equation}
\left<\chi_i^\text{str}| \partial_\theta^2 \chi^\text{str}_i\right> 
=- \sum_{j\neq i} \left|\left<\chi_j^\text{str}| \partial_\theta \chi^\text{str}_i\right> \right|^2
=- \sum \limits_{j\neq i} \left|\frac{\left<\chi_j^\text{str}\left|\left[\frac{\partial}{\partial \theta}, H^\text{str}\right]\right| \chi^\text{str}_i\right>}{E^\text{str}_i(\theta)-E^\text{str}_j(\theta)} \right|^2  \ .
\label{eqn:appendix_secondordercouplings}
\end{equation}
The commutator involving $H^\text{str}$ given in (\ref{eqn:Hamiltonian_strechting}) can be evaluated explicitly to
\begin{equation}
\left[\frac{\partial}{\partial \theta}, H^\text{str}\right]=\left(\frac{\partial} {\partial \theta} H^\text{str}\right)= \sin \theta \frac{\partial}{\partial R_1}\frac{\partial}{\partial R_2} +  \left(\frac{\partial}{\partial \theta} \epsilon(R_1,R_2,\theta)\right)
\end{equation}
and we can discuss the numerator in (\ref{eqn:appendix_secondordercouplings}) in the limit $R^\text{eq} \gg \Delta r$. 
All bending states analyzed in this work localize around equilibrium angles $\theta=0$ or $\theta=\pi$ where $\sin \theta$ vanishes. Going to the limit $R^\text{eq} \gg \Delta r$ increases the moment of inertia and localizes the states so strongly to $\theta=0$ or $\theta=\pi$ that derivative couplings $\sin \theta  \partial_{R_1} \partial_{R_2}$ can be neglected for those states. Furthermore in the limit $R^\text{eq} \gg \Delta r$ we approximate $\left<\chi_j^\text{str}| \left(\partial_\theta \epsilon(R_1,R_2,\theta)\right) |\chi^\text{str}_i\right> \approx  \left(\partial_\theta \epsilon(R^\text{eq},R^\text{eq},\theta) \right) \left<\chi_j^\text{str} |\chi^\text{str}_i\right> =0$ which vanishes due to the orthogonality of the stretching states.

In conclusion this analysis shows, for the model potential strictly and for ULRM under certain assumptions, that the channel equations (\ref{eqn:bend_appendix}) can be decoupled adiabatically in the limit $R^\text{eq} \gg \Delta r$ and reduce to

\begin{equation}
0=\left[ A_{i} + E_i^\text{str}(\theta)-E_\nu \right] \chi_{i\nu}^\text{ben}(\theta) 
\end{equation}

with the operator
\begin{align}
A_{i}&= -\frac{1}{m} \left( \bra{\chi^\text{str}_i} \frac{1}{R_1^2}\ket{\chi^\text{str}_i}+ \bra{\chi^\text{str}_i} \frac{1}{R_2^2}\ket{\chi^\text{str}_i}- \bra{\chi^\text{str}_i} \frac{1}{R_1 R_2}\ket{\chi^\text{str}_i} \cos \theta \right)\left(\frac{\partial^2}{\partial \theta^2} + \cot \theta \frac{\partial}{\partial \theta}  \right)  \nonumber \\
&-\frac{1}{m} \left(\bra{\chi^\text{str}_i} \frac{1}{R_1 R_2}\ket{\chi^\text{str}_i}
\right)\left(\cos \theta + \sin \theta \frac{\partial}{\partial \theta} \right ) \ .
\end{align}

E.g. for the triatomic Rb $43s$ system presented in Fig.\ \ref{fig:benmodes_rb_40} one finds $R_1^\text{eq}/\Delta R_1=R_2^\text{eq}/\Delta R_2\approx 28$. 
%%%%%%%%%%%%%%%%%%%%%%%%%%%%%%%%%%%%%%%%%%%%%%%%%%%%%%%%%%%%%%%%%%%%%%%%%%%%%%%%%%%%%%%%%%%%%  
\end{document}